\documentclass[%
 reprint,
superscriptaddress,
 amsmath,amssymb,
 aps,
 pre,
]{revtex4-1}

\usepackage[utf8]{inputenc}
\usepackage{graphicx}
\usepackage{dcolumn}
\usepackage{bm}
\usepackage{xcolor}
\usepackage[bf]{subfigure}
\usepackage{amsmath}
\usepackage{tcolorbox}

\begin{document}
\title{Contrarian effects and echo chamber formation in opinion dynamics}

\author{Henrique F. de Arruda}
\affiliation{S\~ao Carlos Institute of Physics, University of S\~ao Paulo, S\~ao Carlos, SP, Brazil.}

\author{Alexandre Benatti}
\affiliation{S\~ao Carlos Institute of Physics, University of S\~ao Paulo, S\~ao Carlos, SP, Brazil.}

\author{Filipi Nascimento Silva}
\affiliation{Indiana University Network Science Institute, Indiana University, Bloomington, Indiana 47408, USA.}

\author{C\'esar Henrique Comin}
\affiliation{Department of Computer Science, Federal University of S\~ao Carlos, S\~ao Carlos, Brazil}

\author{Luciano da Fontoura Costa}
\affiliation{S\~ao Carlos Institute of Physics, University of S\~ao Paulo, S\~ao Carlos, SP, Brazil.}

\date{\today}

\begin{abstract}
The relationship between the topology of a network and specific types of dynamics unfolding in networks constitutes a subject of substantial interest.  One type of dynamics that has attracted increasing attention because of its several potential implications is opinion formation.  A phenomenon of particular importance, known to take place in opinion formation, is echo chambers' appearance. In the present work, we approach this phenomenon, while emphasizing the influence of contrarian opinions in a multi-opinion scenario. To define the contrarian opinion, we considered the \emph{Underdog} effect, which is the eventual tendency of people to support the less popular option. We also considered an adaptation of the Sznajd dynamics with the possibility of friendship rewiring, performed on several network models. We analyze the relationship between topology and opinion dynamics by considering two measurements: opinion diversity and network modularity.  Two specific situations have been addressed:  (i) the agents can reconnect only with others sharing the same opinion; and (ii) same as in the previous case, but with the agents reconnecting only within a limited neighborhood. This choice can be justified because, in general, friendship is a transitive property along with subsequent neighborhoods (e.g.,~two friends of a person tend to know each other). As the main results, we found that the Underdog effect, if strong enough, can balance the agents' opinions. On the other hand, this effect decreases the possibilities of echo-chamber formation. We also found that the restricted reconnection case reduced the chances of echo chamber formation and led to smaller echo chambers.
\end{abstract}

\maketitle


\section{\label{sec1} Introduction}
The increasing number of online social network users has impacted several aspects of human interactions and activities, such as  votes in elections~\cite{galam2004contrarian}, opinions about products~\cite{pookulangara2011cultural}, and debates about controversial subjects~\cite{acemouglu2013opinion}. To better understand such phenomena, many aspects of these dynamics have been studied~\cite{gomes2019mobility,gracia2011coevolutionary}, which includes the mechanisms of influence~\cite{sznajd2000opinion} and social perception~\cite{lee2019homophily}. Part of these studies consider the dynamics executed on a network structure, in which the nodes and edges represent people and their friendship, respectively~\cite{sznajd2000opinion,benatti2019opinion}. Other approaches also consider time-varying topologies~\cite{gracia2011coevolutionary,he2004sznajd,holme2006nonequilibrium,fu2008coevolutionary,durrett2012graph,iniguez2009opinion,benatti2019opinion}, where specific rewiring rules are studied.

As human beings are progressively interconnected, several important phenomena have been identified, including the formation of echo chambers~\cite{del2015echo, tornberg2018echo, jasny2015empirical, jasny2018shifting, benatti2019opinion}. More specifically, people sharing the same ideas tend to form relatively isolated communities in social networks. One can define echo chambers on networks as opinions adhered to network communities~\cite{quattrociocchi2016echo,del2016spreading,cinelli2020echo}. Because of its importance, echo chambers have been extensively studied recently~\cite{del2015echo, tornberg2018echo, jasny2015empirical, jasny2018shifting, benatti2019opinion}.

One especially interesting situation deserving further investigation regards networks in which agents can rewire their connections as a consequence of  opinion changes~\cite{he2004sznajd,holme2006nonequilibrium,fu2008coevolutionary,durrett2012graph,iniguez2009opinion,benatti2019opinion}.  In particular, a modified version of the Sznajd model of opinion dynamics was employed~\cite{benatti2019opinion}, considering several network topologies, in order to study echo chamber formation when agents are allowed to reconnect, after changing their opinion, to other agents sharing the new opinion. Furthermore, we considered an arbitrary number of opinions, representing several real scenarios -- for example, elections with many distinct candidates, brand and musical preferences, and among others.

In the present work, we address this problem further with the focus on the effect of contrarian opinions~\cite{dong2018survey,galam2004contrarian,crokidakis2014impact,galam2007role}.  More specifically, when changing their opinion, some people would tend to adopt the position contrary to the predominant opinion. Because, here, we incorporate many options of opinions, we considered the less predominant as being the contrarian. This choice can be explained by the \emph{Underdog} effect, which is the tendency of a part of the public to support the less popular option~\cite{vandello2007appeal,ulmer1978selecting,frazier1991underdog}. What would be the effects of this type of dynamics on the underlying network?  Could this contribute to a broader diversity of opinions and/or promote the echo chamber formation?  

Our proposed dynamics is based on a modified version of the Sznajd model~\cite{benatti2019opinion}. This dynamics considers that the connections relate not only to friendship but also to possible interactions with other people. More specifically, in the real world, users of an online social network can have many friends but typically can communicate effectively only with a small portion of them. To investigate the effects of the contrarians in this context, we include a new rule that allows the individuals to change their opinions to the contrarian, with a given probability. 
We also considered the scenario in which the agents reconnect only within a limited neighborhood, henceforth called \emph{context-based reconnection}. This type of reconnection can be understood as a manner to simulate the fact that a person tends to know the friends of his/her friends~\cite{watts1998collective}.

Since echo chambers can be associated with the adherence of opinions to communities, the comparison between them can quantify the echo chamber formation. In order to compare the agent's opinions with network communities, we employ \emph{modularity}~\cite{newman2006modularity}. More specifically, instead of considering the detected communities, we use the agent's opinions to calculate this measurement. As a complementary analysis, the opinion distribution is also quantified concerning its \emph{diversity}~\cite{jost2006entropy}, which estimates the effective number of opinions.

Several interesting results have been obtained, including the identification of the significant influence of the average degree on the formation of the echo chambers, in both considered situations. In addition, the obtained results were found to exhibit complementary characteristics as far as diversity and modularity are concerned. In particular, we observed that the modularity tended to vary little in regions of the parameter space characterized by similar diversity values, and vice versa. The intensity of the Underdog effect can be associated with the balance of the agent's opinions. More specifically, low and high intensities of Underdog effect can lead the dynamics to echo chamber formation and higher opinion diversity, respectively. Another interesting finding relates to the verification that the context-based reconnections reduced the chances of echo chamber formation, which also tended to be smaller.  We also observed that, for a given set of parameters, two types of topologies could be obtained: with or without echo chambers.

This article is organized as follows.  We start by presenting a previous related work~\cite{benatti2019opinion} on which the current approach builds upon, including the description of the modified Sznajd dynamics, the reconnecting schemes, the definition of diversity and modularity, as well as the adopted network models. The results are then presented and discussed, and prospects for future studies are suggested.

\section{Contrarian-Driven Sznajd Model}
Before starting the dynamics, each network node, $i$, is assigned to a categorical opinion, $O_i \in \mathbb{N}$, randomly distributed with uniform distribution, where $O_i \in [0,N_O]$. The cases $O_i = 0$ and  $O_i > 0$ corresponds to nodes with null and not-null opinions, respectively. The null opinion means that the individual does not have an opinion about the subject. For instance, in the case of election opinions, the individuals who do not know the candidates hold a null opinion about them.

An additional probability, $w$ ($0 \leq w \leq 1$), can also be employed, which corresponds to the probability of a node randomly changing its opinion. To simplify our analysis, we henceforth adopt $w=0$.

\begin{figure*}
    \centering
    \begin{tcolorbox}
    
    \begin{flushleft}
    A node, $i$, is randomly chosen, then:
    \begin{itemize}
        \item if $ O_i = 0$ the iteration ends;
        \item if $ O_i \neq 0$, a random neighbor of $i$, referred as $j$, is selected. By considering the opinion $O_j$, the next action is determined as:
        \begin{itemize}
            \item if $ O_j = 0$, the node $j$ changes its opinion to agree with the node $i$ ($ O_j = O_i$);
            \item if $ O_j \neq O_i$, the iteration ends;
            \item if $ O_j = O_i$:
            \begin{itemize}
                \item {Each neighbor of $i$ can change its opinions to $O_i$, with probability $1/k_i$. For each neighbor that does not change its opinion, the neighbor can change to the contrarian opinion with probability $g$;}
                \item {Each neighbor of $j$ can change its opinion to $O_i$, with probability $1/k_j$.}
            \end{itemize}
    
            \item When the opinion $O_l$, of given node $l$, changes, one of the following three rules is applied with probability $q$.
            \begin{itemize}
                \item If the new opinion $O_l$ is unique on the network, nothing happens;
                \item If all neighbors of $l$ agree with the new opinion, nothing happens;
                \item If the above two rules are not applied, $l$ loses a connection with an aleatory neighbor that has a different opinion, and connects to some other random node having the same opinion as $l$.
            \end{itemize}
         \end{itemize}
    \end{itemize}
    \end{flushleft}
    
    \end{tcolorbox}
    
    \caption{Pseudocode of the proposed survey-driven Sznajd model. The bold text depicts the differences between the survey-driven Sznajd model and ASM.}
    \label{fig:code2}
\end{figure*}

\begin{figure*}[!htpb]
  \centering
     \includegraphics[width=0.62\textwidth]{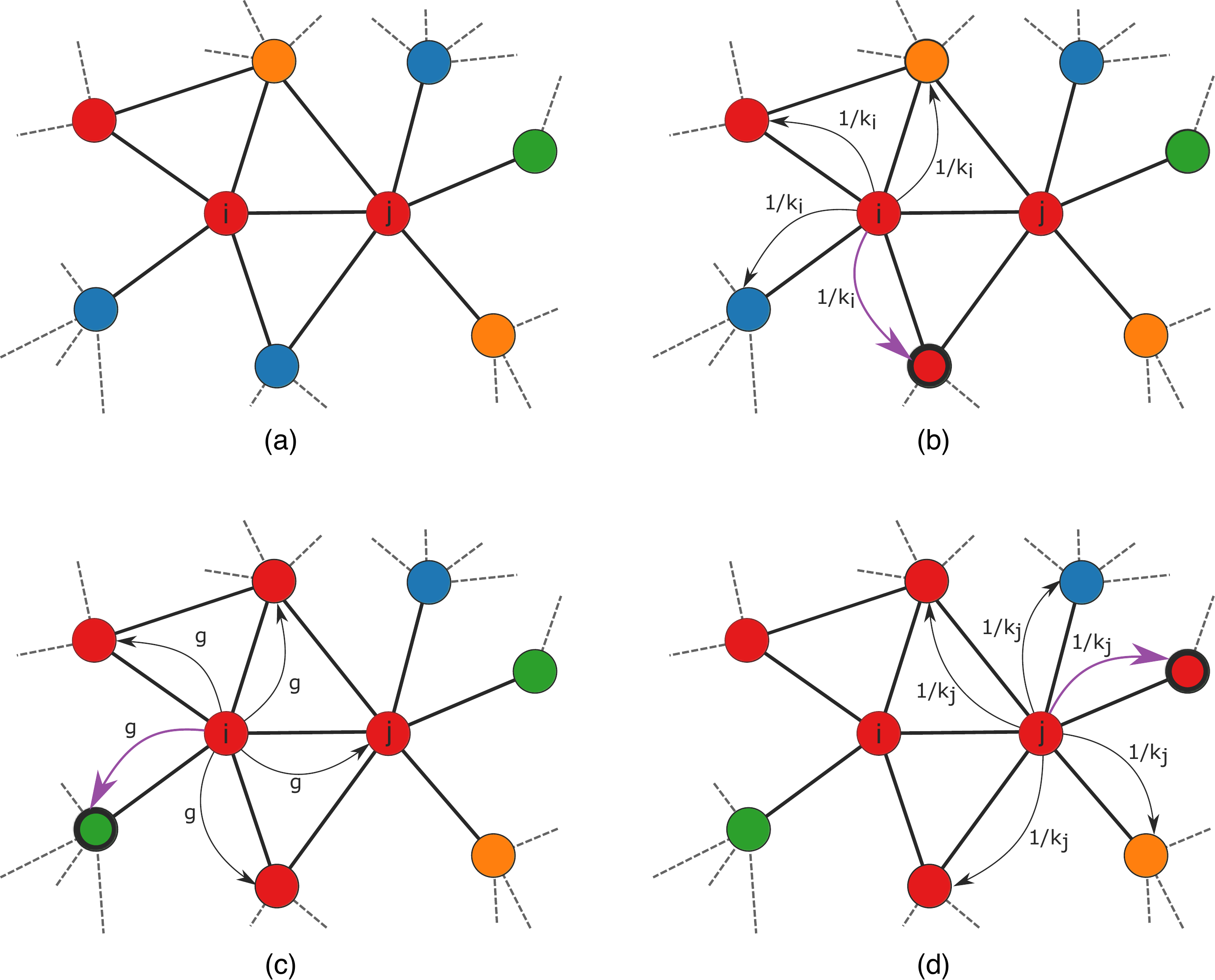}
   \caption{Representation of the contrarian-related processing. (a): Initial configuration example of nodes $i$ and $j$ and their respective neighbors. (b): the probabilities of node $i$ to influence its neighbors, in which the highlighted neighbor was convinced to the opinion of $i$. In this example, only one neighbor had its opinion changed, but all nodes had the same probability of adhering to the opinion of $i$. The Underdog effect is illustrated in (c), given that the green node corresponds to the less frequent opinion. (d): the probabilities of $j$ to convince its neighbors. The highlighted  illustrates the change to opinion $i$.}
  \label{fig:figcode}
\end{figure*}

In order to incorporate the contrarian dynamics, we propose some complementary rules, as presented in Figure~\ref{fig:code2}.
Our proposed dynamics simulates the case in which people influenced by their neighbors can adopt the contrary opinion. Contrarian opinions have been studied by many researchers~\cite{galam2004contrarian,dong2018survey,crokidakis2014impact,galam2007role}. Here, we put together the concepts of persuasion, given by the Sznajd model~\cite{sznajd2000opinion}, change of friendship, and the contrarian effect. In this case, two agents try to convince another, which can change its opinion to the contrarian, adopting the opinion defined according to the Underdog effect. An example of the contrarian processing is illustrated in Figure~\ref{fig:figcode}.

In order to study this effect, we considered the ASM and added a rule that incorporates the contrarian idea, which consists of allowing an agent to have an opinion that is different from the majoritarian. 
More specifically, for each iteration, all agents' opinions are analyzed, and the contrarian opinion is defined as the less frequent one.
Differently from the previous study~\cite{benatti2019opinion}, here we considered the starting number of opinions as four (this choice is better discussed in Section~\ref{sec:variedNumber}). 
Our model incorporates more than two categories of opinions since, for many real cases, it cannot be represented by binary assumptions. For instance, in the case of the earth shape believe, there are at least four possibilities: spherical, flatten, toroidal, and potato shape.

\section{Context-based reconnection}
We also investigate a variation of the Contrarian-Driven Sznajd Model in which the rewirings can be done only between topologically close agents. We incorporated this new rule in the above-described algorithm.  More specifically, we included a parameter $h$, which controls the maximum topological distance between $i$ and $j$, allowing a change of opinion by $i$. So, we limit the reconnections to happen only between nodes that are within a distance lower or equal to $h$. If there is no possibility of reconnection, the rewiring does not happen. For the sake of simplicity, here we adopt $h=2$, which means that the reconnections happen only between the selected node $i$ and the friends of friends of $i$. 

\section{Diversity}
To quantify how diverse the opinions are, we employ a respective measurement. There are many possible ways to define diversity~\cite{jost2006entropy}. Here we consider the variation based on information theory~\cite{pielou1966shannon}, which is defined as follows
\begin{equation}
    D = \exp{(H)},
\end{equation}
where $H$ is the Shannon entropy, which is defined as
\begin{equation}
    H = -\sum_{o=0}^{N_o}{\rho_o \ln(\rho_o)},
\end{equation}
where $N_o$ is the number of possible opinions and $\rho_o$ is the proportion of the opinion $o$ on network. The value of diversity, limited within the range $1 \leq D \leq N_o + 1$, can be understood as the effective number of states, also known as Hill number of order $q=1$~\cite{hill1973diversity,chao2016phylogenetic}. This variation of diversity have been employed to quantify other opinion-based dynamics~\cite{messias2018can,benatti2019opinion}.

\section{Modularity}
Because diversity only accounts for the variety of opinions, we also consider a measurement regarding the topology. We employ the modularity~\cite{newman2006modularity} that quantifies the tendency of nodes to form communities. These communities are defined as groups of nodes highly interconnected while being weakly linked to the remaining network~\cite{newman2006modularity}.

The adopted modularity measurement is calculated as
\begin{equation}
    Q = \dfrac{1}{2m}\sum_{ij}{\Big[A_{ij} - \dfrac{k_i k_j}{2m}\Big]\delta(c_i,c_j)},
\end{equation}
where $m$ is the number of edges, $A$ is the adjacency matrix, and $c_i$, $c_j$ are the communities of the nodes $i$ and $j$, respectively. The value of modularity gauges the structures of clusters of a network.  In this study, we did not detect communities.  Instead, we understand sets of nodes having the same opinion as constituting a respective community.

\section{Network topologies}
To account for different network topologies and to incorporate distinct real-based characteristics, we perform the dynamics considering five different models as follows: 
\begin{itemize}
    \item Watts-Strogatz (WS)~\cite{watts1998WS}: we considered the network created from a 2D toroidal lattice;
    \item Erdős-Rényi (ER)~\cite{erdos1959random}: having
    uniformly random connections with probability $p$;
    \item Barabási–Albert (BA)~\cite{barabasi1999BA}: yielding power-law degree distribution;
    \item Random geometric graph (GEO)~\cite{penrose2003random}: the positions of the nodes were initially distributed on a 2D surface;
    \item Stochastic Block Model (SBM)~\cite{holland1983stochastic}: we configured the model concerning four well-defined communities with the same size.
\end{itemize}

In all the above cases, the parameters were chosen to yield the same expected average degree, $\langle k \rangle$. For all these adopted networks, we considered the number of nodes as being approximately $1000$. Furthermore, we employed three different average degrees ($\langle k \rangle = 4, 8, 12$). However, in the case of the GEO model, we considered only $\langle k \rangle = 8, 12$ since it is difficult to achieve a single connected component with a lower average degree. More information regarding several of the adopted network models can be found in~\cite{costa2007characterization}.

\section{\label{sec:met} Results and discussion}
In this section, we present the results according to two respective subsections considering the no-reconnection constraint and context-based reconnections. In both cases, we analyze the diversity and modularity of the opinions. 

\subsection{No-reconnection constraint}
\begin{figure*}[!htbp]
\centering
\begin{subfigure}[($\langle k \rangle =$~4).]{
\includegraphics[width=0.3\textwidth]{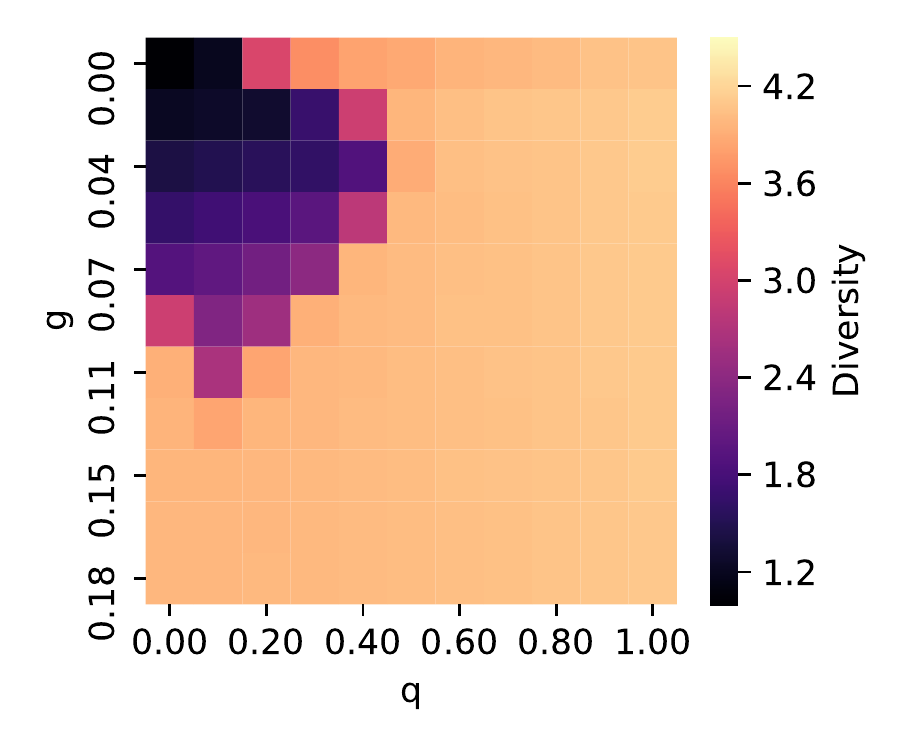}}
\end{subfigure}
~
\begin{subfigure}[($\langle k \rangle =$~8).]{
\includegraphics[width=0.3\textwidth]{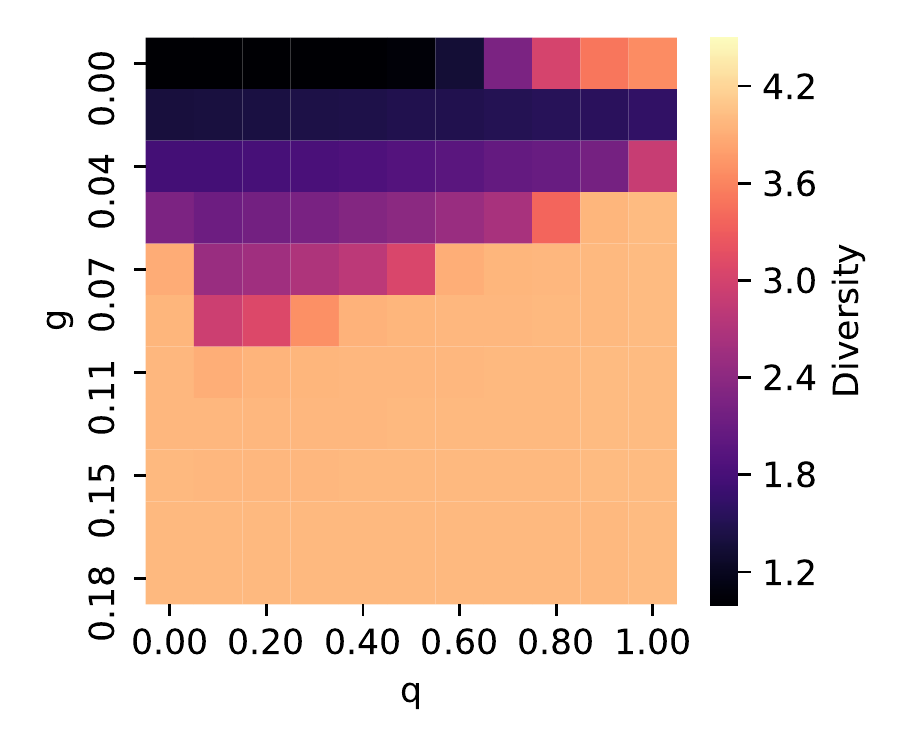}}
\end{subfigure}
~
\begin{subfigure}[($\langle k \rangle =$~12).]{
\includegraphics[width=0.3\textwidth]{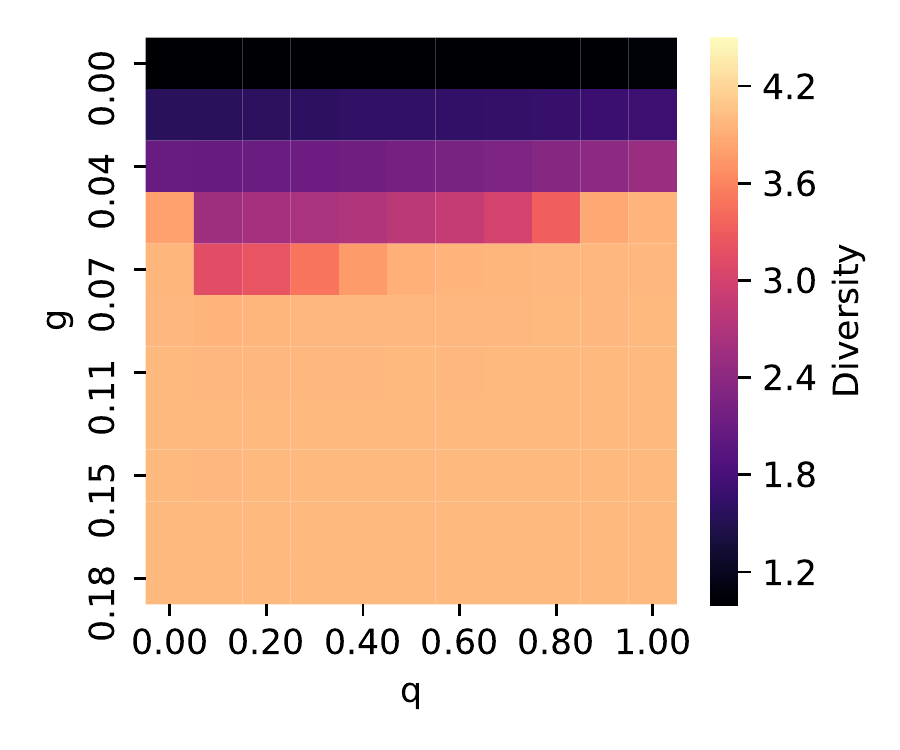}}
\end{subfigure}

\begin{subfigure}[($\langle k \rangle =$~4).]{
\includegraphics[width=0.3\textwidth]{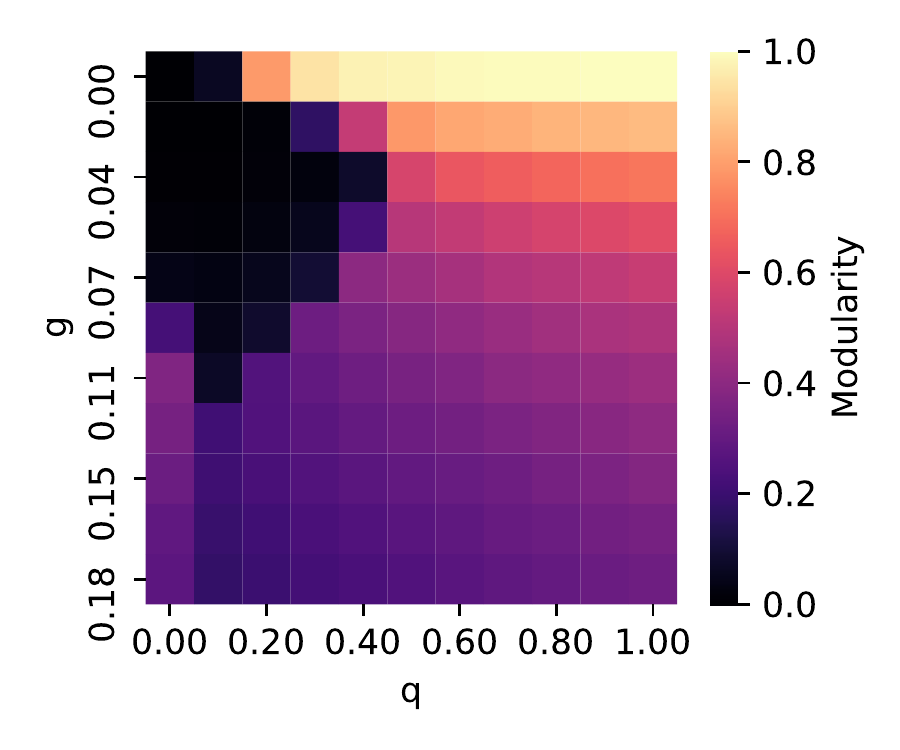}}
\end{subfigure}
~
\begin{subfigure}[($\langle k \rangle =$~8).]{
\includegraphics[width=0.3\textwidth]{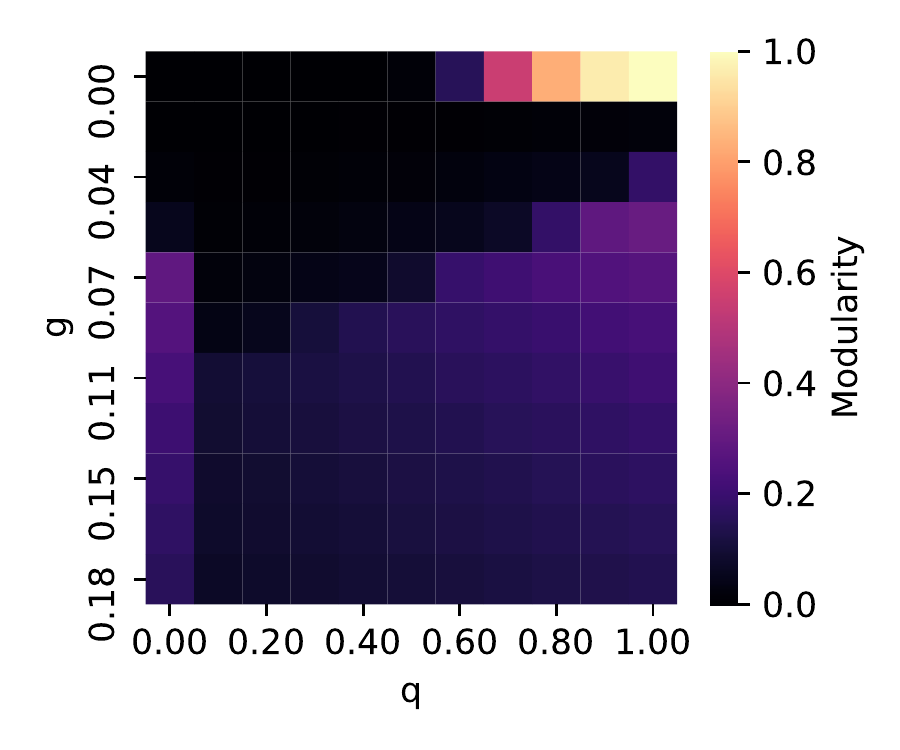}}
\end{subfigure}
~
\begin{subfigure}[($\langle k \rangle =$~12).]{
\includegraphics[width=0.3\textwidth]{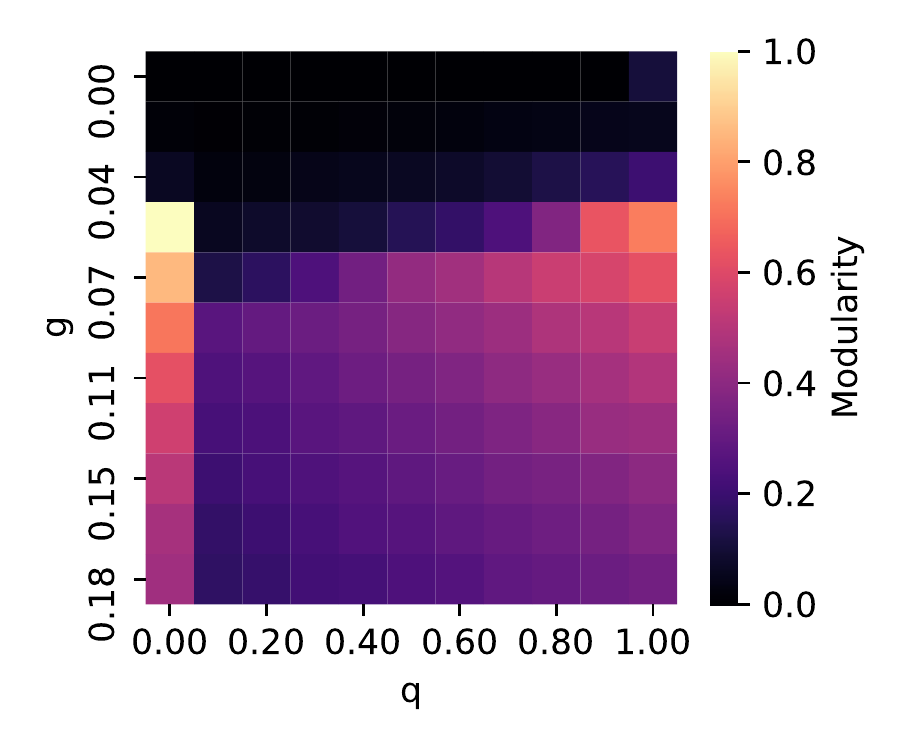}}
\end{subfigure}
\caption{Comparison between $D$ and $Q$, the latter normalized according to the highest value, for a given set of parameters, in which items (a), (b), and (c) are respective to $D$, while (d), (e), and (f) relate to $Q$. The WS network was considered in this example. The variation of $Q$ for $\langle k \rangle = 12$ is much lower due to the high values of average degrees. Each of the computed points was calculated for 100 network samples.}
\label{fig:ws}
\end{figure*}

First, we analyzed the diversity ($D$) behavior in terms of the reconnection probability ($q$) and contrarian probability ($g$) for all considered topologies and three average degrees ($\langle k \rangle = 4,8,12$). For most of the dynamics, we executed 1,000,000 iterations, except for GEO, which was performed 100 million (for average degree 8) and 25 million (for average degree 12). These numbers of iterations were chosen to allow the dynamics to reach a steady state. Furthermore, we repeated each experiment 100 times. For all of the considered topologies, we calculated the average values of $D$ by varying $q$ and $g$. An example regarding the WS network is shown in Figures~\ref{fig:ws}(a)~(b)~and~(c), in which well-defined regions can be observed. For almost all network models, the results were found to be similar. The lower diversity values were observed for lower values of $q$ and $g$. Interestingly, even when we consider $q=0$ (no reconnections), some values of $g$ lead the dynamics to converge to high values of opinion diversity, $D$. In other words, we verified that the employed parameter configuration strongly affects the measurement of diversity ($D$). 

\begin{figure}[!htpb]
  \centering
     \includegraphics[width=0.48\textwidth]{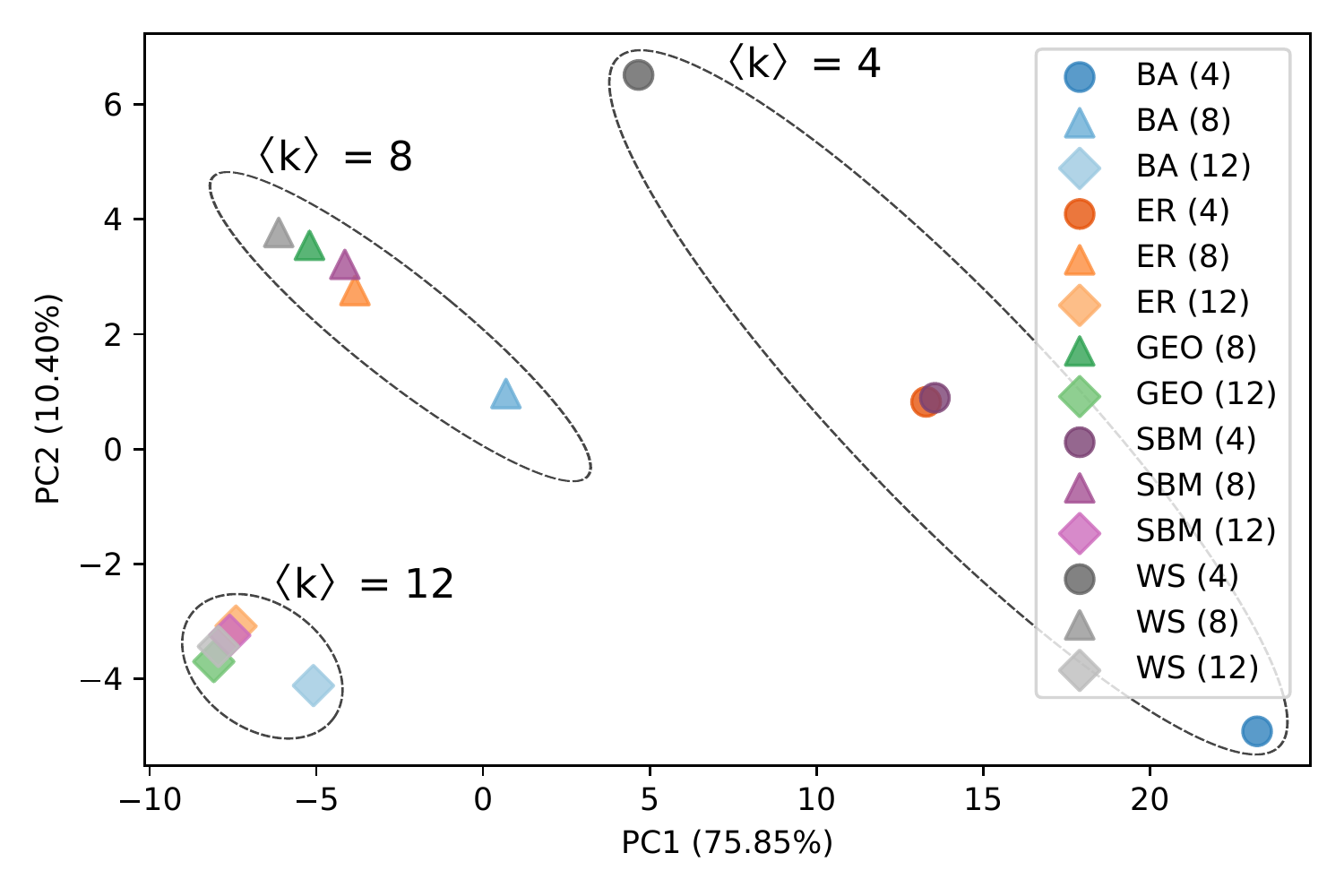}
   \caption{PCA projection of $D$, by employing the same set of parameters as Figure~\ref{fig:ws}. It is possible to observe groups of samples (identified by ellipses), according to $\langle k \rangle$.}
  \label{fig:pca1}
\end{figure}

In order to better understand the variation of the diversity with the parameters, we flattened the obtained values of $D$ and calculated the respective PCA (Principal Component Analysis) projection~\cite{jolliffe2011principal,gewers2018principal} (see Figure~\ref{fig:pca1}). More specifically, for each network type, we compute a matrix of $D$ similar to the results shown in Figure~\ref{fig:ws}. Also, we flattened the obtained values of $D$ and, by considering these vectors, we obtained the PCA projection. An interesting result concerns the separation of the cases into three regions in terms of the average degree, identified by respective ellipses in Figure~\ref{fig:pca1}.  For the two highest values of average degree, the samples were found to be more tightly clustered. Furthermore, for $\langle k \rangle = 4$, the group is more widely scattered.  This result suggests that the average degree plays a particularly important role in defining the characteristics of the opinion dynamics in the considered cases.

Next, we analyzed how the opinions modularity ($Q$) changes according to the model parameters. Because average degrees can influence the network modularity~\cite{fortunato2010community}, for all of the matrices, we divide the values by the highest average value. This procedure was not employed to $D$ because, in this case, the obtained result is related to the effective number of opinions. We compute $Q$ for all network variations, and for $\langle k \rangle = 4, 8, 12$ using the same set of parameters we employed in the previous case. Figures~\ref{fig:ws}(d)~(e)~and~(f) illustrate examples of $Q$ for WS networks, in which well-defined regions can also be found. For the highest of the considered values of $g$ and $q$, high values of $Q$ were obtained, except for $\langle k \rangle = 12$. It means that there is a possibility to have echo chambers. The other parameter configurations led to networks without well-defined communities. Similar results were also observed for the other models. For higher average degrees, $Q$ tends to be lower for all possibilities of parameters ($g$ and $q$).  Another critical aspect involved in interpreting the $Q$ measurement is setting the limit of detection~\cite{fortunato2007resolution}. For example, in the cases in which $D>4$, there are disconnected nodes that have a null opinion.

Now, we proceed to discuss the results obtained for diversity and modularity in an integrated way. The modularity analysis reveals a pattern not evidenced by the diversity analysis (see Figure~\ref{fig:ws}). For the highest values of diversity $D$, the modularity $Q$ was found to be more sensitive to parameter variations. More specifically, while $D$ is found to measure almost always similar values in this region, $Q$ displays a broader variation. In a complementary fashion, for the lowest values of modularity, the diversity was found to be more sensitive (as can be seen in Figure~\ref{fig:ws}). In general, both measurements are equally important to describe the presented dynamics behavior. Furthermore, the formation of the echo chamber can happen only for high values of $D$ and $Q$. In other words, $D$ describes the effective number of opinions, and $Q$ is a quantification of the community organization.

\begin{figure*}[!htpb]
  \centering
     \includegraphics[width=0.65\textwidth]{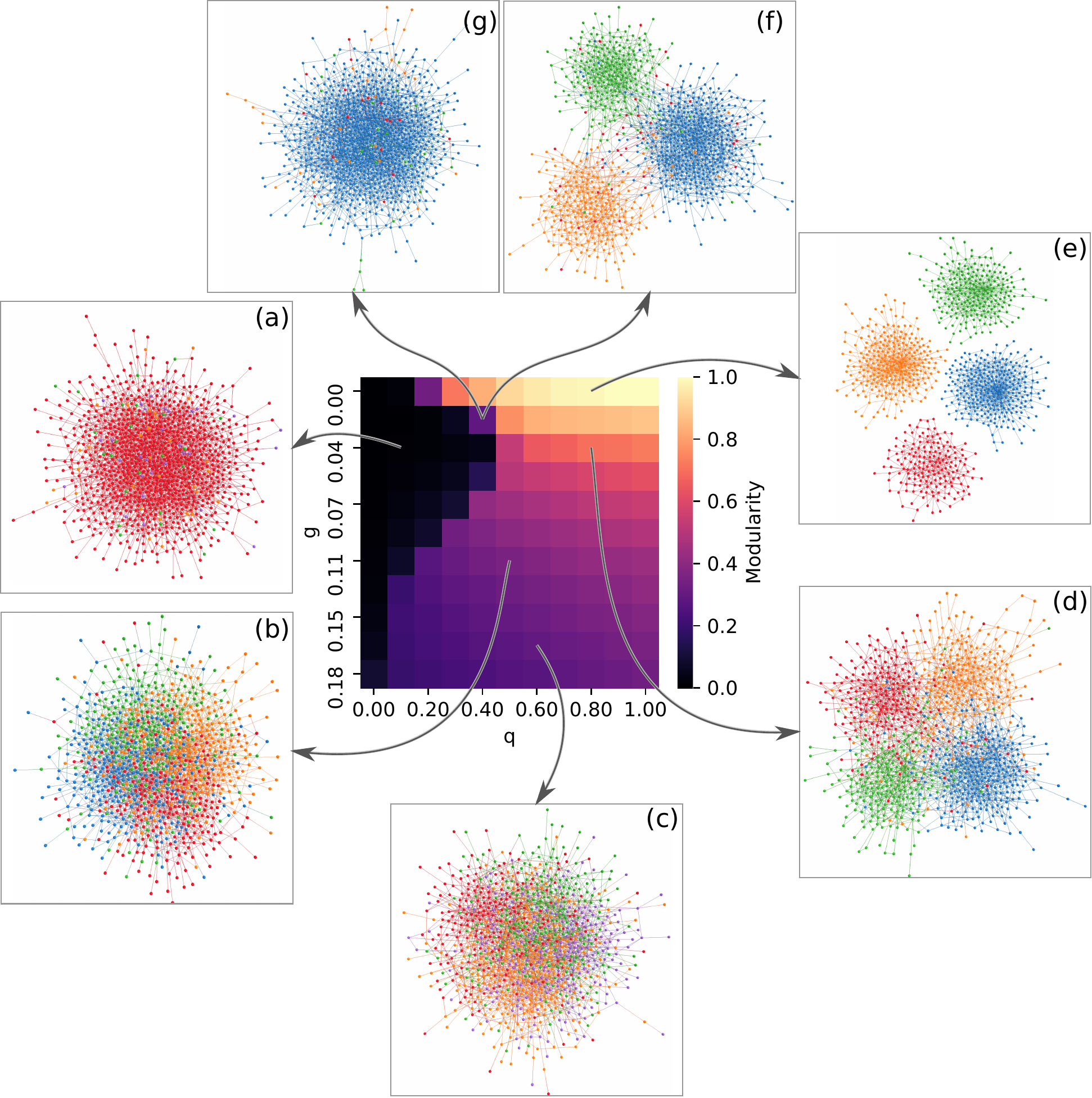}
   \caption{Some examples of the resulting networks for given parameters. The heatmap represents $Q$ values (normalized to have maximum value equals to one) obtained after the execution of our dynamics. Here, we employ the BA network, for $\langle k \rangle = 4$. Interestingly, for $q=0.40$ and $g=0.02$ more than one type of network organization can be obtained. The node colors in the network visualizations represent the opinions. Each of the computed points was calculated for 100 network samples. The network visualizations were created using the software implemented in~\cite{silva2016using}.}
  \label{fig:figNetworkExamples}
\end{figure*}

Figure~\ref{fig:figNetworkExamples} illustrates the resulting topologies when starting with BA networks. More specifically, we present a heatmap of $Q$ values and some respective examples of the resulting networks. In the well-defined region with $Q$ next to zero, the dynamics converge to a single opinion (see Figure~\ref{fig:figNetworkExamples}(a)).  Figures~\ref{fig:figNetworkExamples}(b)~and~(c) were obtained in regions with intermediate values of $Q$. In this case, the communities are not well-defined. Even so, in both cases, there is a high level of diversity, indicated by the visualization colors. Networks with distinct communities were obtained for large $q$ and $g$ -- see  Figures~\ref{fig:figNetworkExamples}(d)~and~(e). Thus, larger reconnection and contrarian probabilities further the formation of echo chambers. The network shown in Figure~\ref{fig:figNetworkExamples}(e) has communities that are disconnected among themselves. For some configurations, both behaviors, with and without community structure, can be found for the same parameter configurations (see Figures~\ref{fig:figNetworkExamples}(f)~and~(g)). The value displayed in the matrix means an average, where for Figure~\ref{fig:figNetworkExamples}(f), $Q$ is much higher than for Figures~\ref{fig:figNetworkExamples}(g), in which $Q$ is found to be near to zero. Figures~\ref{fig:figNetworkExamples}(f)~and~(g)). Interestingly, this situation was also identified for another opinion dynamics (ASM), reported in~\cite{benatti2019opinion}.

\subsection{Context-based reconnection}
In this subsection, we explore the effects of the proposed dynamics when the interactions are restricted. This constraint simulates the fact that people tend to become a friend of a friend ($h=2$). In this case, we considered only the SBM and GEO networks because these networks have higher diameters than the other considered models. So, the effect of the context-based reconnection is more visible. 

\begin{figure*}[!htbp]
\begin{subfigure}[SBM~($\langle k \rangle =$~4).]{
\includegraphics[width=0.3\textwidth]{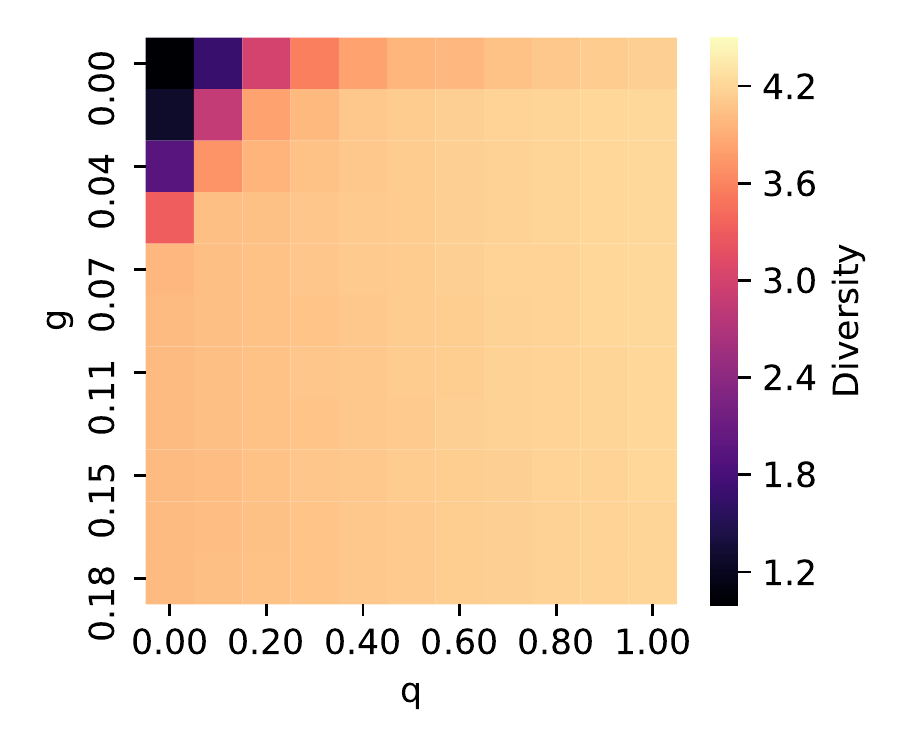}}
\end{subfigure}
~
\begin{subfigure}[SBM~($\langle k \rangle =$~8).]{
\includegraphics[width=0.3\textwidth]{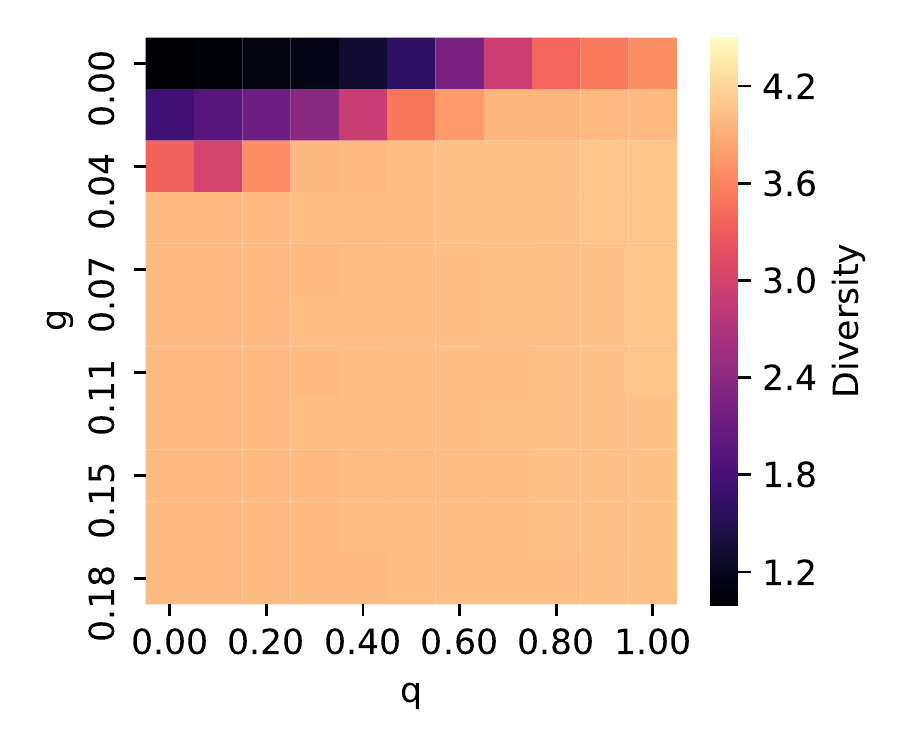}}
\end{subfigure}
~
\begin{subfigure}[SBM~($\langle k \rangle =$~12).]{
\includegraphics[width=0.3\textwidth]{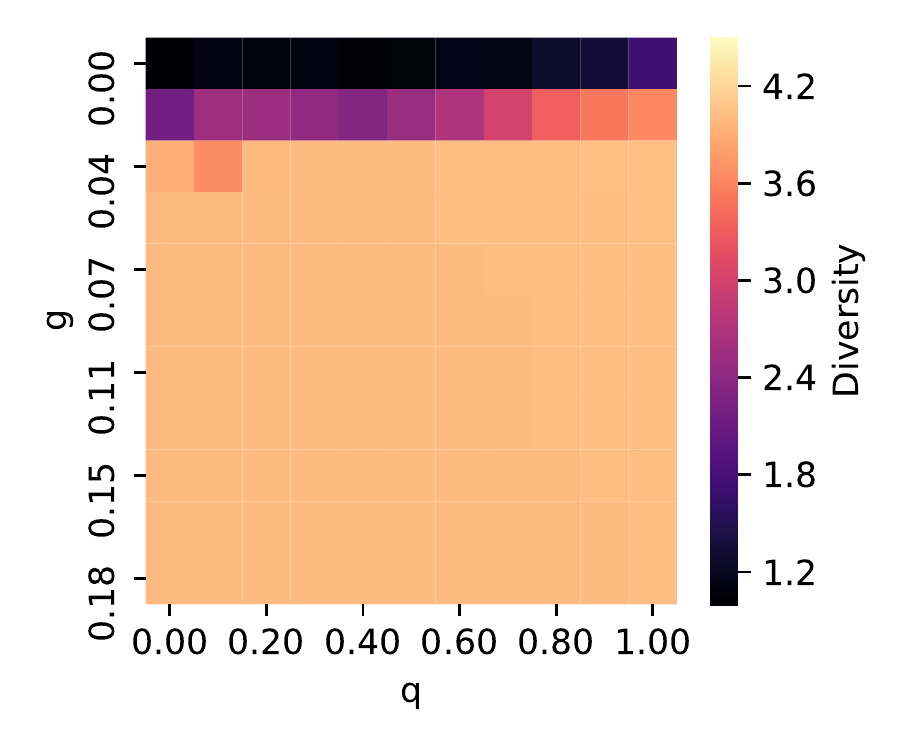}}
\end{subfigure}

\begin{subfigure}[SBM~($\langle k \rangle =$~4).]{
\includegraphics[width=0.3\textwidth]{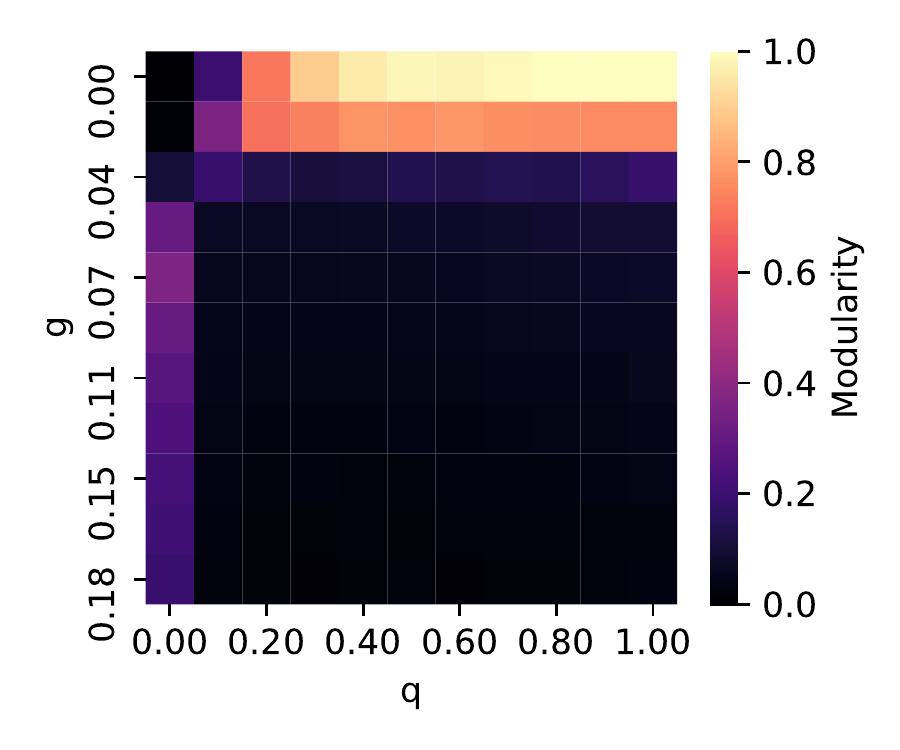}}
\end{subfigure}
~
\begin{subfigure}[SBM~($\langle k \rangle =$~8).]{
\includegraphics[width=0.3\textwidth]{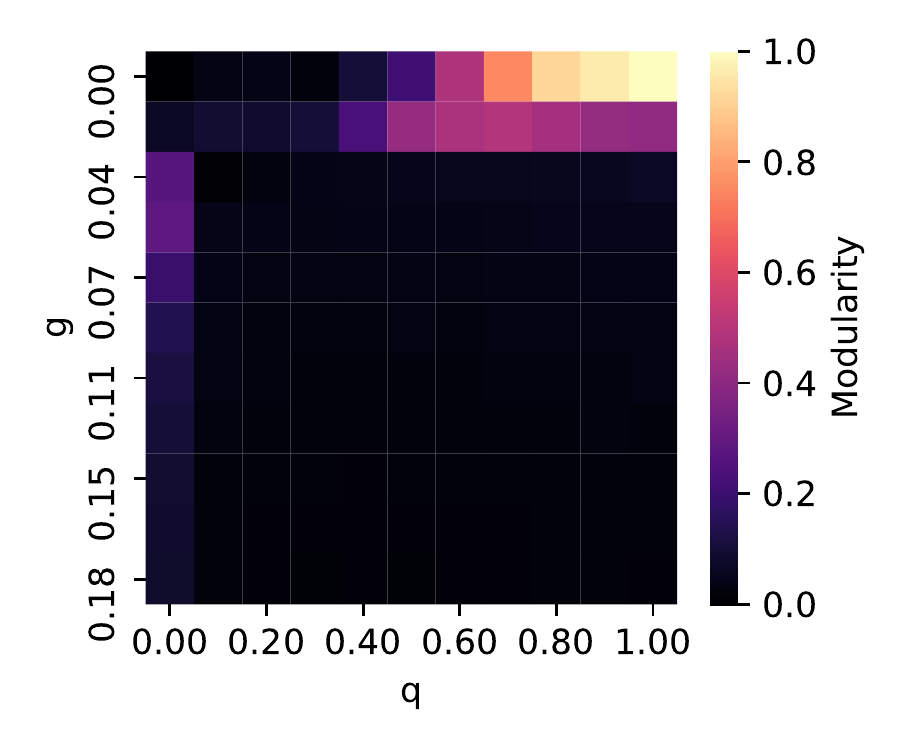}}
\end{subfigure}
~
\begin{subfigure}[SBM~($\langle k \rangle =$~12).]{
\includegraphics[width=0.3\textwidth]{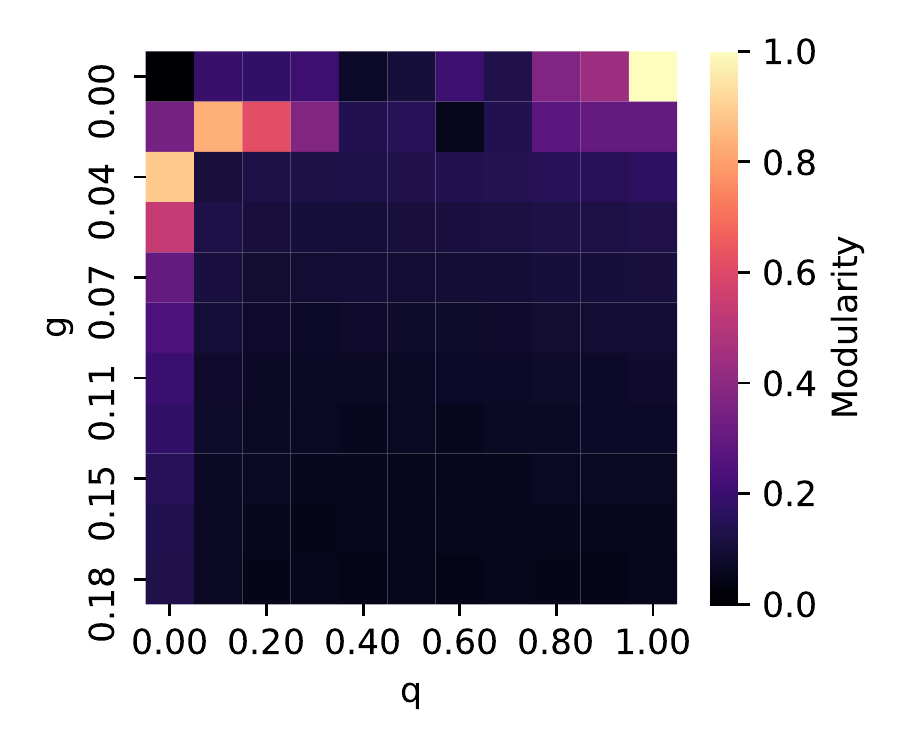}}
\end{subfigure}

\caption{Comparison between $D$ and $Q$, the latter normalized according to the highest value, for a given set of parameters, in which items (a), (b), and (c) are respective to $D$, while (d), (e), and (f) relate to $Q$. Here, we considered SBM networks and the context-based reconnection dynamics ($h=2$).}
\label{fig:sbm_h}
\end{figure*}

By considering the diversity ($D$), the results were found to be similar to the no-rewiring constraint dynamics (see Figure~\ref{fig:sbm_h}(a)~(b)~and~(c)). However, the regions with lower values of $D$ are found only for smaller regions defined by specific combinations of parameters. Also, comparing with the previous model, the modularity values were found to be different. In this case, the region in which $Q$ tends to zero is considerably ampler.

\begin{figure*}[!htbp]
\centering
\begin{subfigure}[$q=0.00$ and $g=0.09$]{
\includegraphics[width=0.3\textwidth]{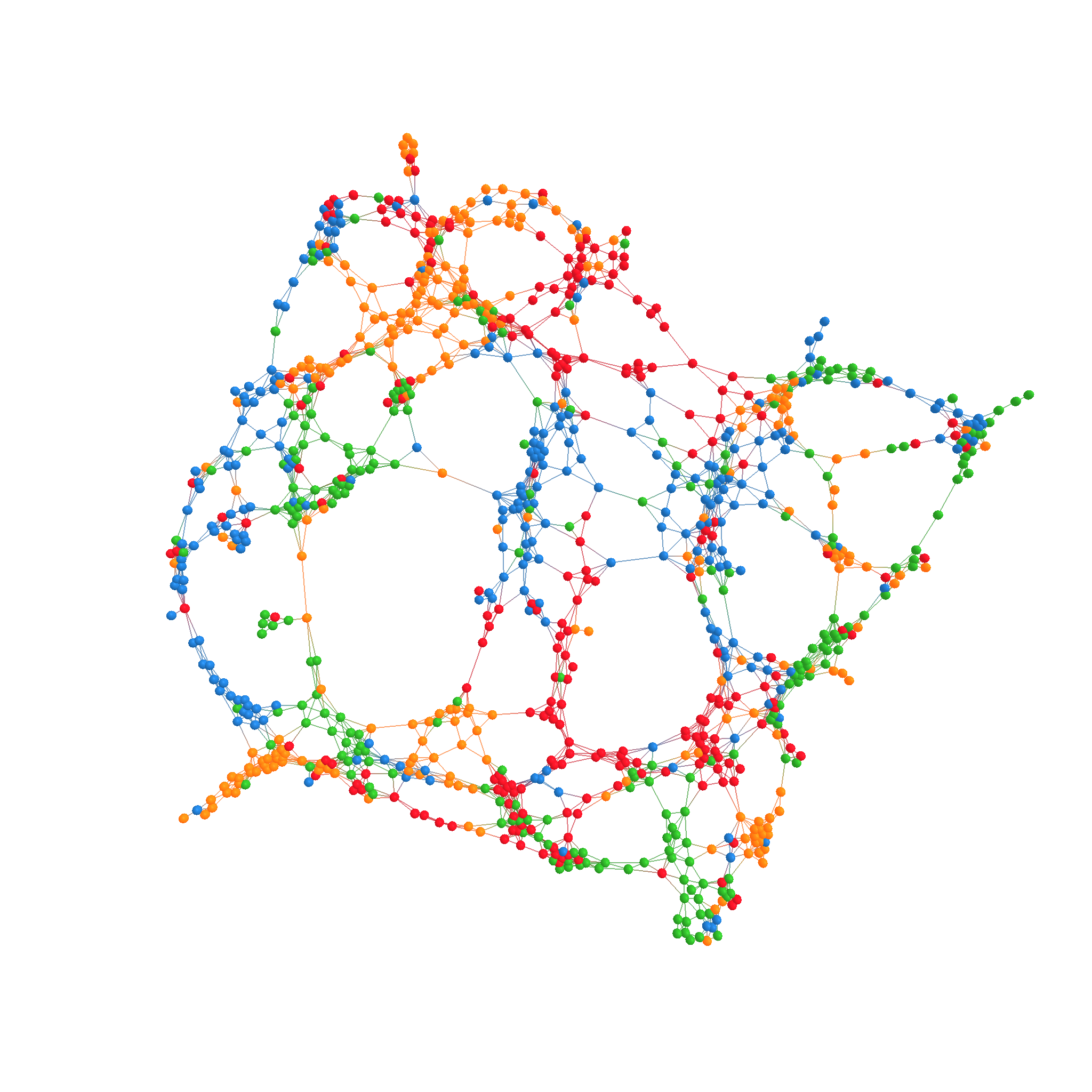}}
\end{subfigure}
~
\begin{subfigure}[$q=0.10$ and $g=0.02$]{
\includegraphics[width=0.3\textwidth]{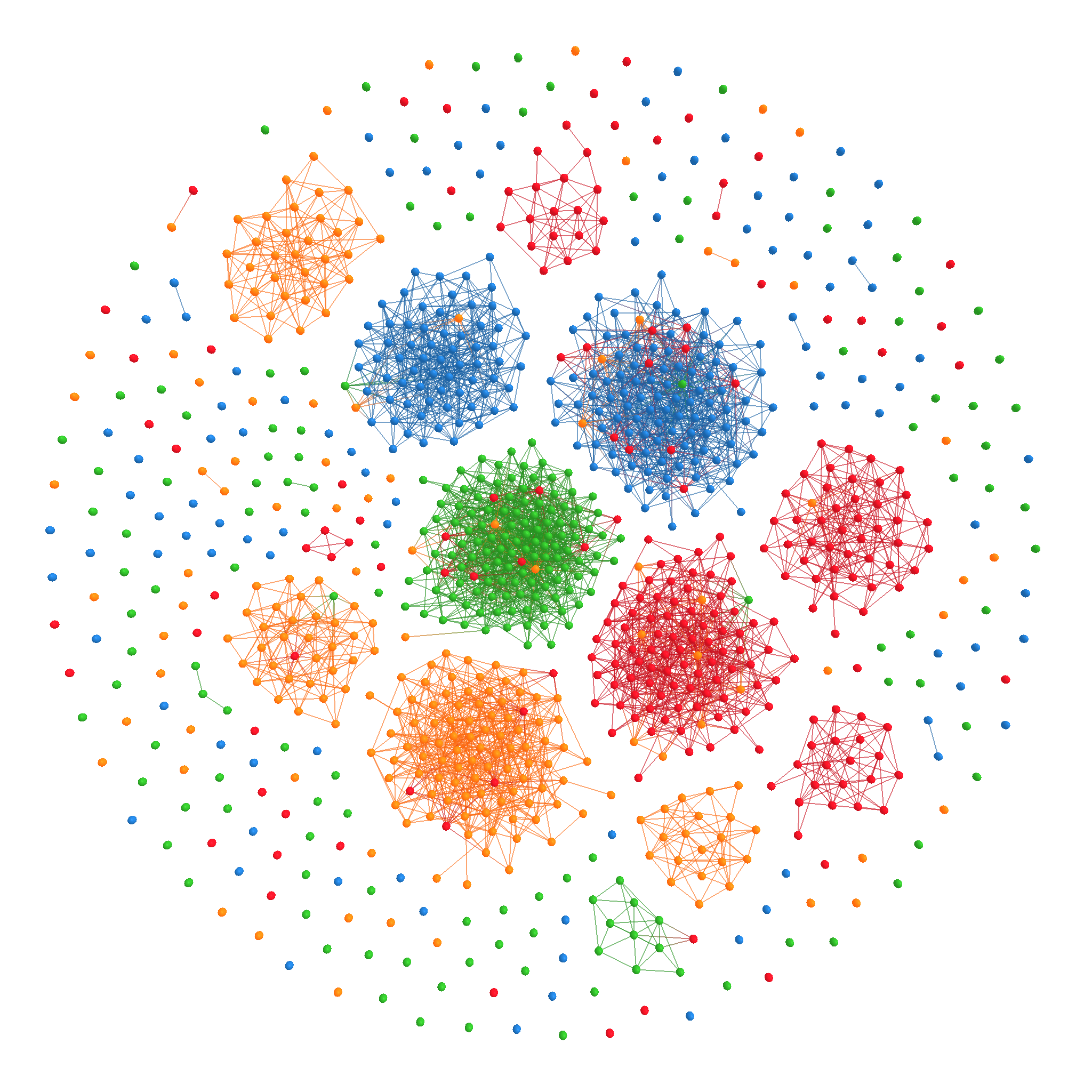}}
\end{subfigure}
~
\begin{subfigure}[$q=0.30$ and $g=0.09$]{
\includegraphics[width=0.3\textwidth]{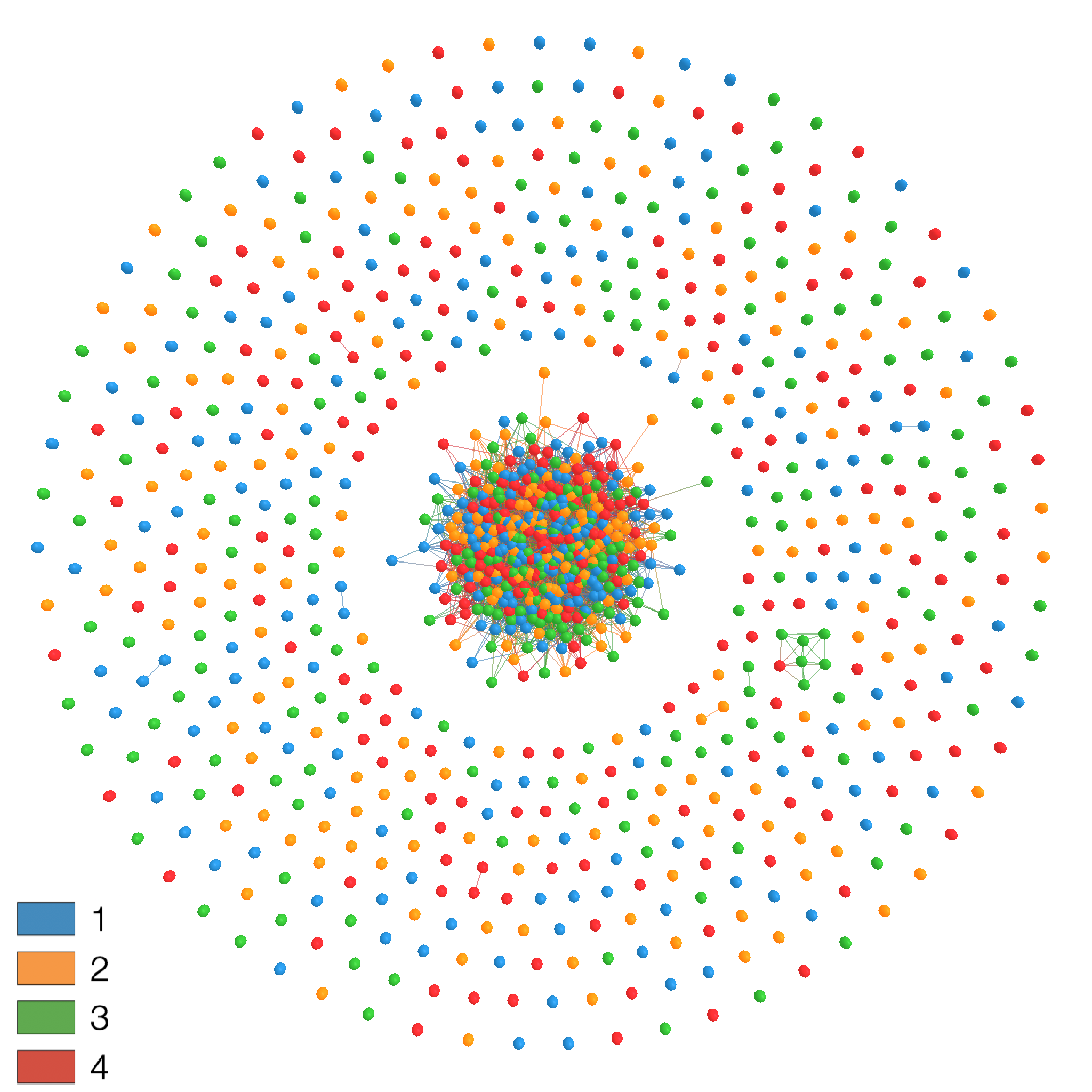}}
\end{subfigure}

\caption{Visualizations of resultant topologies when starting with GEO networks ($\langle k \rangle$ = 8). The employed dynamics is based on the context-based reconnection ($h=2$). Each of the computed points was calculated for 100 network samples. These network visualizations were created using the software implemented in~\cite{silva2016using}.}
\label{fig:geo_nets_h}
\end{figure*}

Figure~\ref{fig:geo_nets_h} shows some possible resulting topologies when starting with GEO networks ($\langle k \rangle$ = 8). Figure~\ref{fig:geo_nets_h}(a) illustrates an example for $q=0$ (no reconnections are allowed), characterized by high value of $D$ and low value of $Q$. The opinions were found to define relatively small groups. In the case of Figure~\ref{fig:geo_nets_h}(b), there is also a wide range of opinions, but with the formation of echo chambers. Furthermore, nodes from completely separated communities can have the same opinion. Figure~\ref{fig:geo_nets_h}(c) shows another possibility of resulting network with high value of $D$ and low value of $Q$. As in the previous result, isolated nodes can also be found. In summary, by considering this restriction ($h=2$), we found that it is much easier to have parameters that give rise to high diversity. However, high modularity is observed only within a more restricted region defined by $g$ and $q$.  

\subsection{Varied numbers of opinions}
\label{sec:variedNumber}

\begin{figure*}[!htbp]
\begin{subfigure}[$N_O = 2$]{
\includegraphics[width=0.3\textwidth]{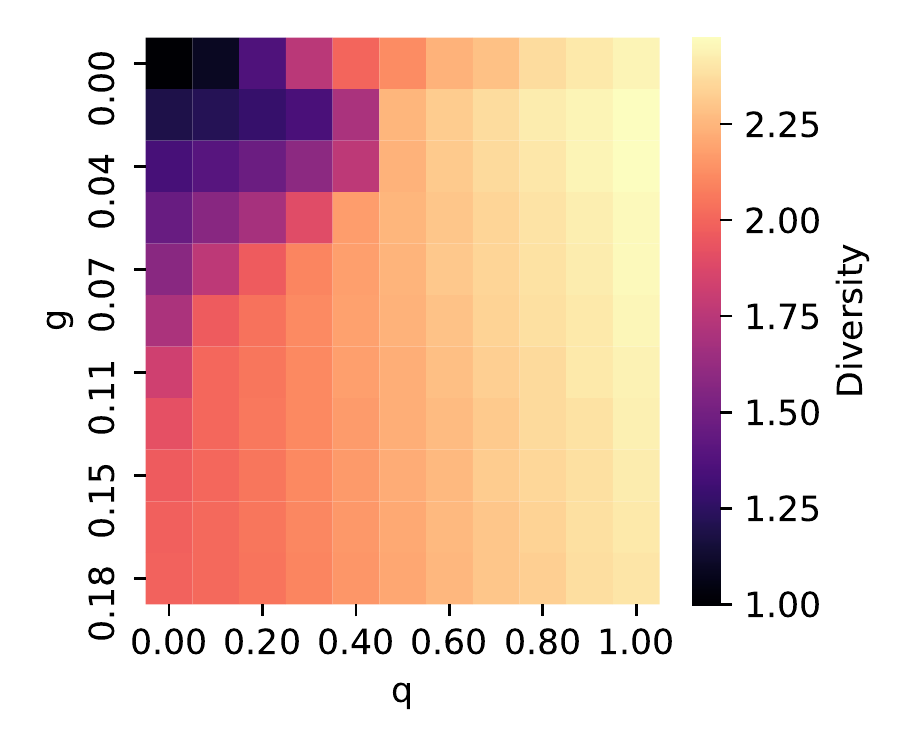}}
\end{subfigure}
~
\begin{subfigure}[$N_O = 3$]{
\includegraphics[width=0.3\textwidth]{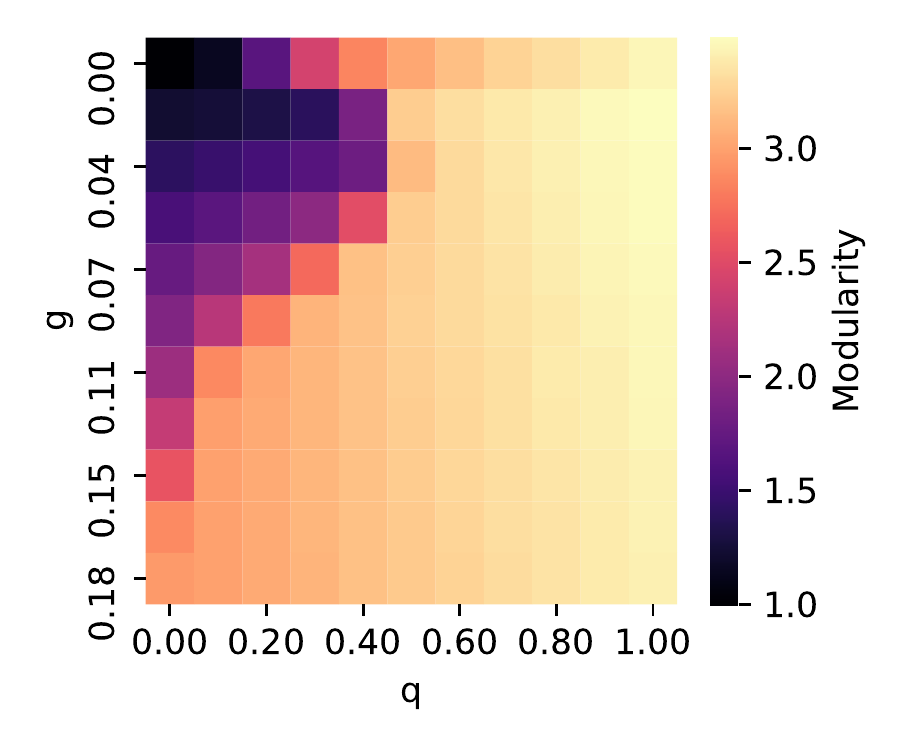}}
\end{subfigure}
~
\begin{subfigure}[$N_O = 4$]{
\includegraphics[width=0.3\textwidth]{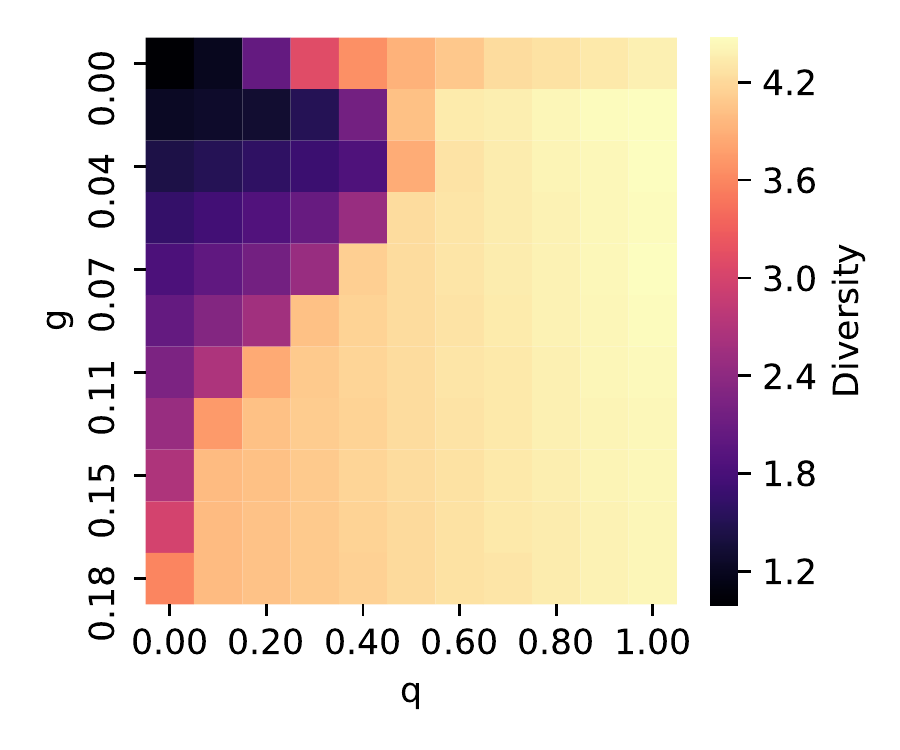}}
\end{subfigure}

\begin{subfigure}[$N_O = 5$]{
\includegraphics[width=0.3\textwidth]{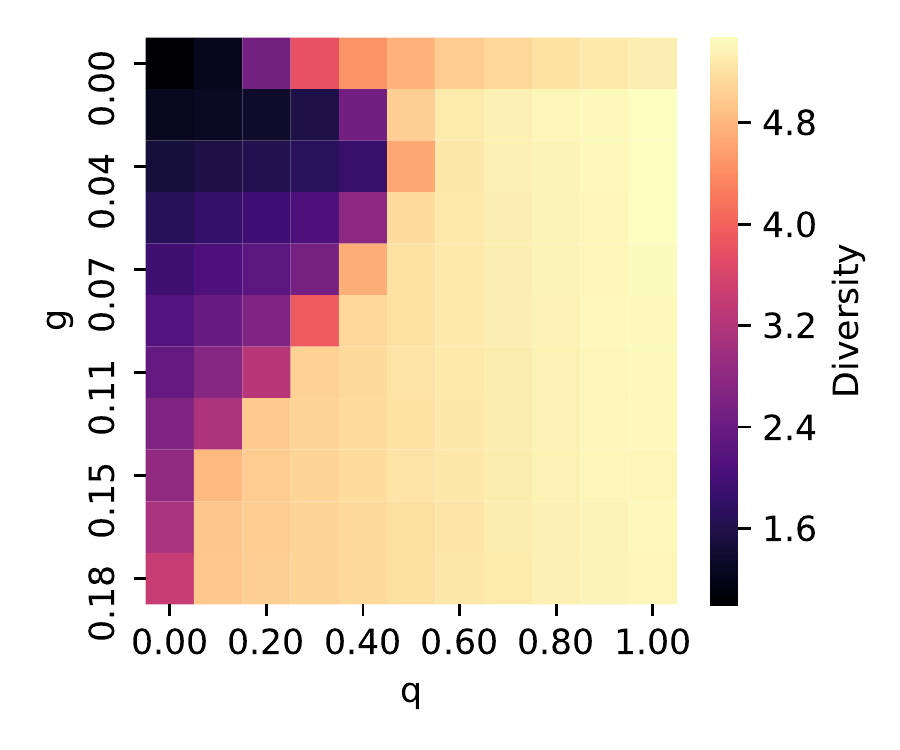}}
\end{subfigure}
~
\begin{subfigure}[$N_O = 6$]{
\includegraphics[width=0.3\textwidth]{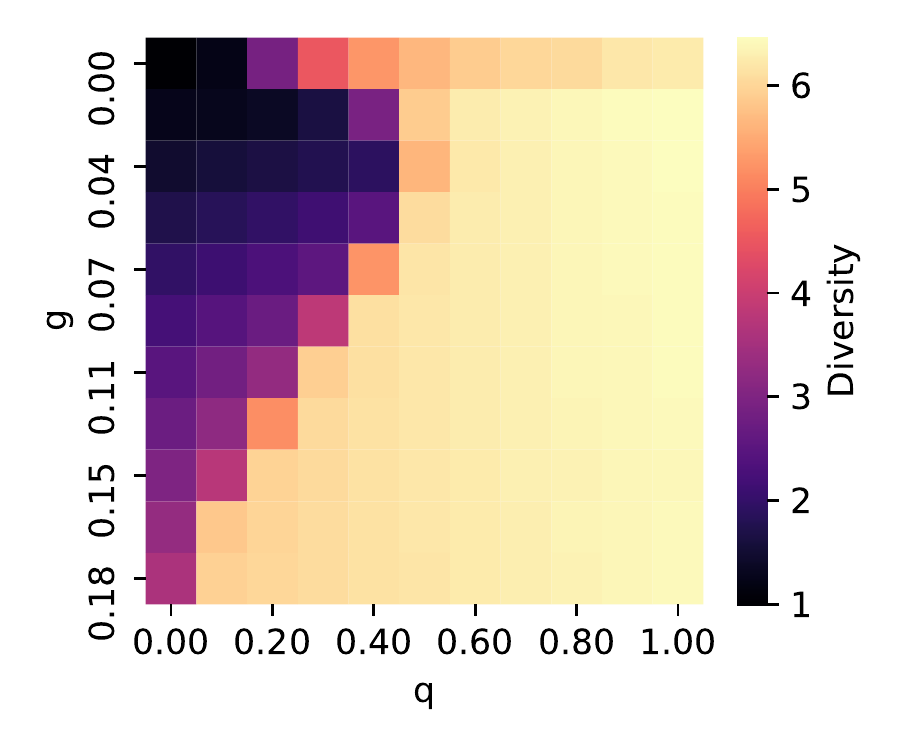}}
\end{subfigure}

\caption{$D$ obtained for BA networks (with $\langle k \rangle$ = 4), with varied numbers of opinions.}
\label{fig:diversity_n_communities}
\end{figure*}

\begin{figure*}[!htbp]
\begin{subfigure}[$N_O = 2$]{
\includegraphics[width=0.3\textwidth]{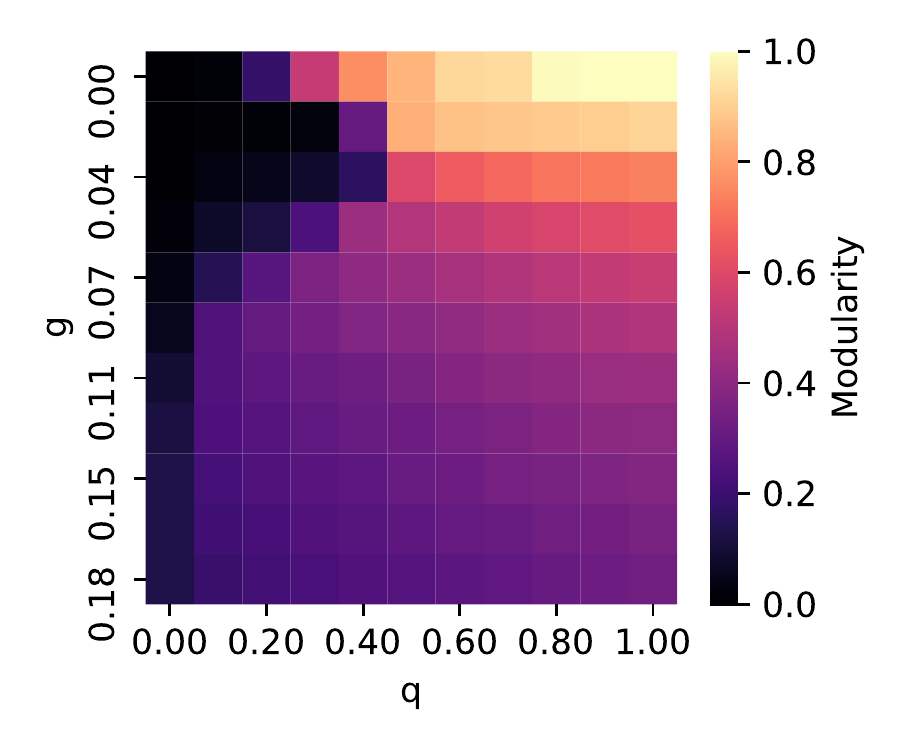}}
\end{subfigure}
~
\begin{subfigure}[$N_O = 3$]{
\includegraphics[width=0.3\textwidth]{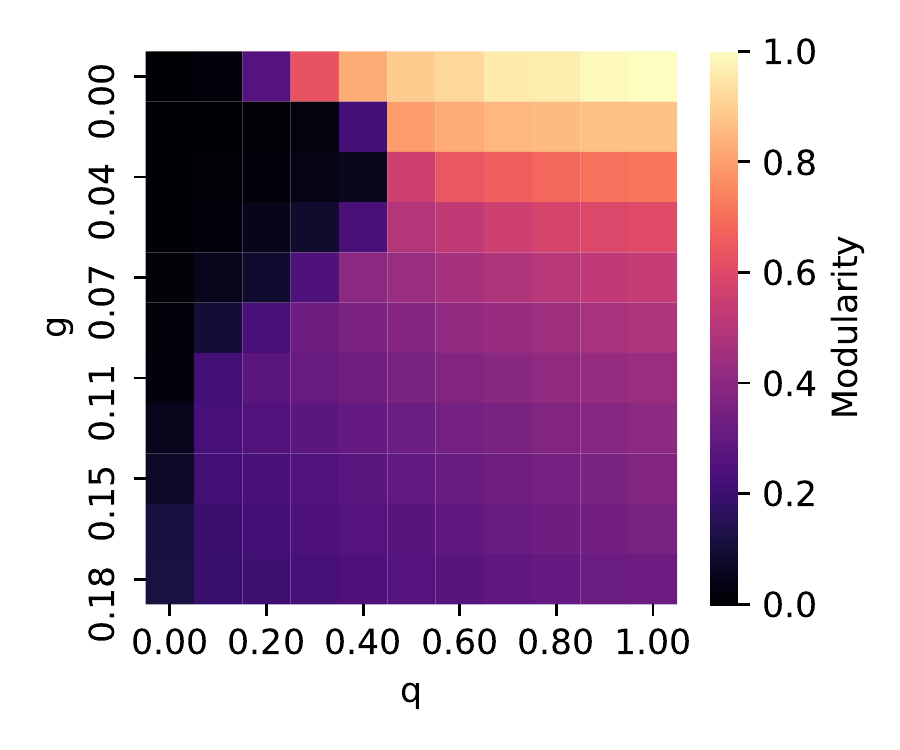}}
\end{subfigure}
~
\begin{subfigure}[$N_O = 4$]{
\includegraphics[width=0.3\textwidth]{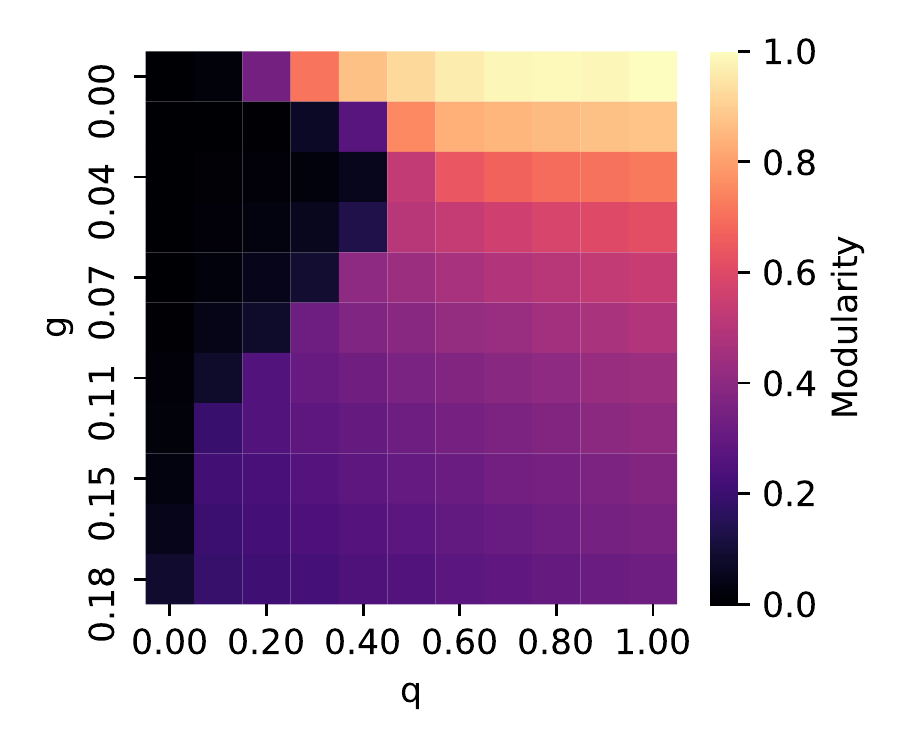}}
\end{subfigure}
~
\begin{subfigure}[$N_O = 5$]{
\includegraphics[width=0.3\textwidth]{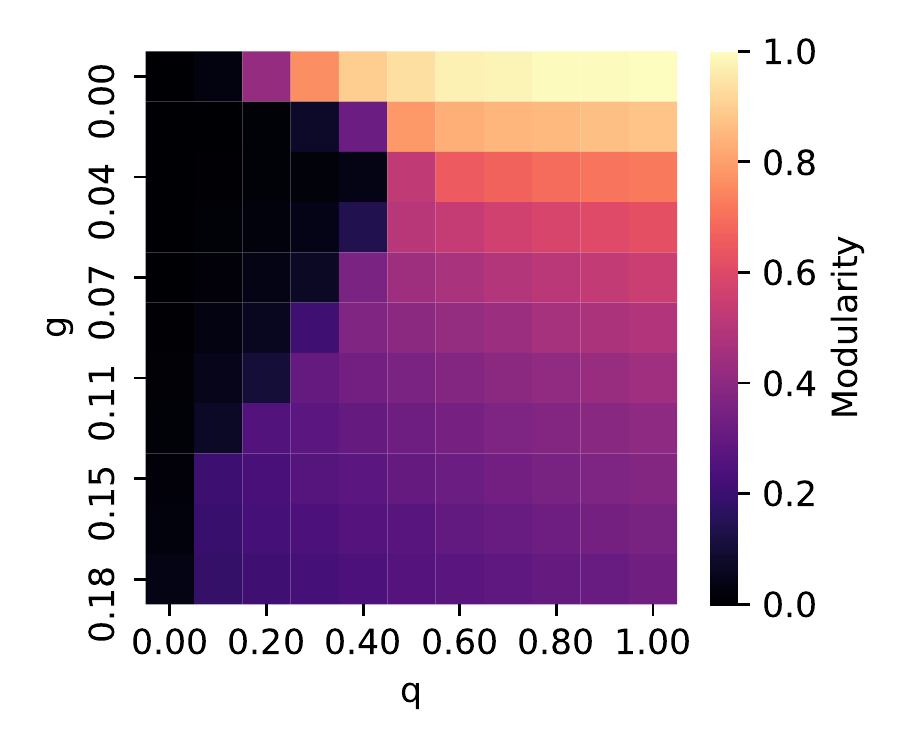}}
\end{subfigure}
~
\begin{subfigure}[$N_O = 6$]{
\includegraphics[width=0.3\textwidth]{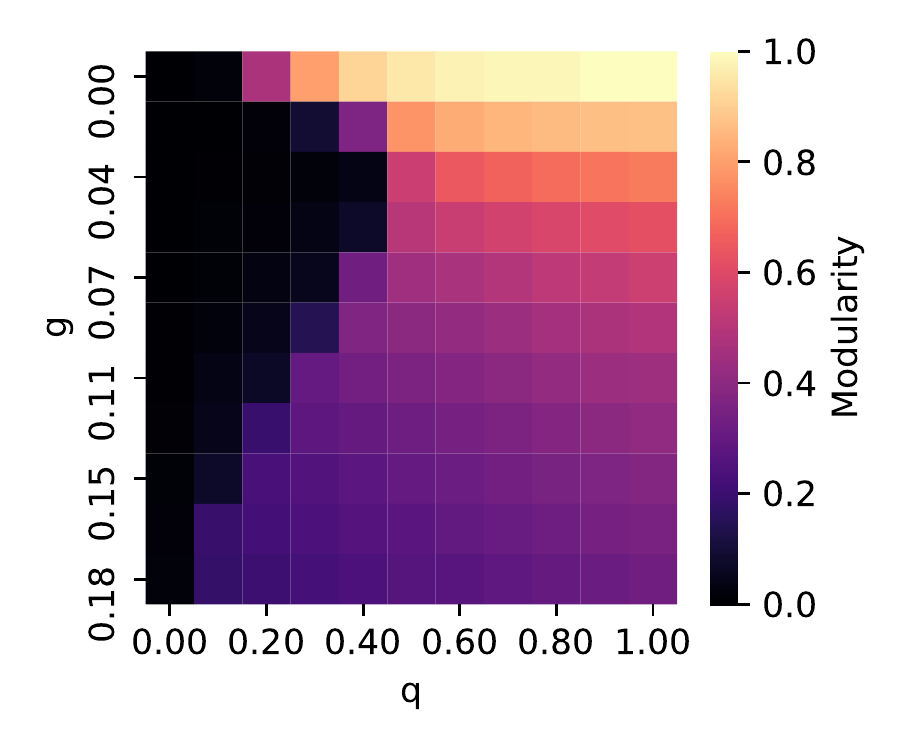}}
\end{subfigure}

\caption{Normalized version of $Q$ obtained for BA networks, with $\langle k \rangle$ = 4, witch varied numbers of opinions.}
\label{fig:modularity_n_communities}
\end{figure*}

In this subsection, we compare the execution of the dynamics by varying the number of opinions, $N_O$ (2, 3, 4, 5, and 6 opinions). Here, we considered a single network topology, which is the same topology we considered in Figure~\ref{fig:figNetworkExamples}. More specifically, we employ BA networks with $\langle k \rangle = 4$. Figures~\ref{fig:diversity_n_communities}~and~\ref{fig:modularity_n_communities} illustrates the obtained results of $D$ and $Q$, respectively. All in all, the results are found to be similar. For all of the tested cases, the number of both scenarios, of a single or $N_O$ opinions, were found. Additionally, the measured values of $Q$ mean that the dynamics drove the to have both types of topology, with or without communities. In the case of $D$, the set of parameters that give rise to the lowest values (Figure~\ref{fig:diversity_n_communities}) are ordered in increasing order according to $N_O$. Similar results were found for $Q$ (Figure~\ref{fig:modularity_n_communities}), where $N_O$ is also related to the sets of parameters that result in networks without communities.

\section{\label{sec:conc} Conclusions}
Several studies have addressed the topic of opinion formation, and in particular, echo chamber formation in social networks. In the present work, we approached the problem of echo chamber formation in several types of complex networks, as modeled by a modified Sznajd model.  In particular, we focused attention on the effects of contrarian opinions by considering a time-varying topology. More specifically, we incorporate the Underdog effect, and multiple opinion possibilities. Furthermore, two strategies have been tested, which include: (i) the agents can be reconnected only with others sharing the same opinion, and (ii) the agents also can be reconnected only with other that share the same opinion, but within a limited neighborhood.

Several interesting results have been obtained and discussed. Regarding the analysis based on diversity and modularity, the obtained results were found to exhibit complementary characteristics. More specifically, we found that some regions of the parameter space are characterized by a gradual variation of diversity while displaying very similar modularities, and vice versa. For specific parameter configurations, two types of topologies can be observed: with or without the presence of echo chambers. In addition, by considering both types of dynamics, the contrarians, conducted by the Underdog effect, play an important role in the resultant topology and dynamics. In the case of high contrarian probability, $g$, the effective number of opinions tends to be high. However, when low values of $g$ are employed, less well-defined communities are found. In other words, the Underdog effect can contribute to an increase in the heterogeneity of opinions. In our result, this effect can as well balance the group sizes.

Moreover, one of the factors that strongly influences the dynamics was found to be the average degree, which is particularly determinant on the formation of the echo chambers. This result means that the average number of friends plays an important role in the dynamics. In the case of the context-based reconnections, it reduced the chances of echo chamber formation, which also tended to be smaller. By considering the number of opinions, we found that this parameter did not strongly affect the steady-state of the dynamics. However, the formed echo chambers reflect the number of opinions. Observe also that in this work, it was decided not to analyze our dynamics analytically because our model is particularly intricate as a consequence of many possible parameter configurations.

The findings reported in this article motivate several further investigations.  In particular, it would be interesting to study the effect of the spontaneous opinion changes. Also, continuous variables could be adopted in order to characterize opinions. Another possibility is to consider weighted and/or directed complex networks.

\section*{Acknowledgments}
Henrique F. de Arruda acknowledges FAPESP for sponsorship (grant no. 2018/10489-0 and no. 2019/16223-5). Alexandre Benatti thanks Coordenação de Aperfeiçoamento de Pessoal de N\'ivel Superior - Brasil (CAPES) - Finance Code 001. Luciano da F. Costa thanks CNPq (grant no. 307085/2018-0) and NAP-PRP-USP for sponsorship. C\'esar H. Comin thanks FAPESP (grant number no. 18/09125-4) for sponsorship. This work has been supported also by FAPESP grants no. 11/50761-2 and no. 2015/22308-2.

\bibliography{ref}

\begin{thebibliography}{44}%
\makeatletter
\providecommand \@ifxundefined [1]{%
 \@ifx{#1\undefined}
}%
\providecommand \@ifnum [1]{%
 \ifnum #1\expandafter \@firstoftwo
 \else \expandafter \@secondoftwo
 \fi
}%
\providecommand \@ifx [1]{%
 \ifx #1\expandafter \@firstoftwo
 \else \expandafter \@secondoftwo
 \fi
}%
\providecommand \natexlab [1]{#1}%
\providecommand \enquote  [1]{``#1''}%
\providecommand \bibnamefont  [1]{#1}%
\providecommand \bibfnamefont [1]{#1}%
\providecommand \citenamefont [1]{#1}%
\providecommand \href@noop [0]{\@secondoftwo}%
\providecommand \href [0]{\begingroup \@sanitize@url \@href}%
\providecommand \@href[1]{\@@startlink{#1}\@@href}%
\providecommand \@@href[1]{\endgroup#1\@@endlink}%
\providecommand \@sanitize@url [0]{\catcode `\\12\catcode `\$12\catcode
  `\&12\catcode `\#12\catcode `\^12\catcode `\_12\catcode `\%12\relax}%
\providecommand \@@startlink[1]{}%
\providecommand \@@endlink[0]{}%
\providecommand \url  [0]{\begingroup\@sanitize@url \@url }%
\providecommand \@url [1]{\endgroup\@href {#1}{\urlprefix }}%
\providecommand \urlprefix  [0]{URL }%
\providecommand \Eprint [0]{\href }%
\providecommand \doibase [0]{http://dx.doi.org/}%
\providecommand \selectlanguage [0]{\@gobble}%
\providecommand \bibinfo  [0]{\@secondoftwo}%
\providecommand \bibfield  [0]{\@secondoftwo}%
\providecommand \translation [1]{[#1]}%
\providecommand \BibitemOpen [0]{}%
\providecommand \bibitemStop [0]{}%
\providecommand \bibitemNoStop [0]{.\EOS\space}%
\providecommand \EOS [0]{\spacefactor3000\relax}%
\providecommand \BibitemShut  [1]{\csname bibitem#1\endcsname}%
\let\auto@bib@innerbib\@empty
\bibitem [{\citenamefont {Galam}(2004)}]{galam2004contrarian}%
  \BibitemOpen
  \bibfield  {author} {\bibinfo {author} {\bibfnamefont {S.}~\bibnamefont
  {Galam}},\ }\href@noop {} {\bibfield  {journal} {\bibinfo  {journal} {Physica
  A: Statistical Mechanics and its Applications}\ }\textbf {\bibinfo {volume}
  {333}},\ \bibinfo {pages} {453} (\bibinfo {year} {2004})}\BibitemShut
  {NoStop}%
\bibitem [{\citenamefont {Pookulangara}\ and\ \citenamefont
  {Koesler}(2011)}]{pookulangara2011cultural}%
  \BibitemOpen
  \bibfield  {author} {\bibinfo {author} {\bibfnamefont {S.}~\bibnamefont
  {Pookulangara}}\ and\ \bibinfo {author} {\bibfnamefont {K.}~\bibnamefont
  {Koesler}},\ }\href@noop {} {\bibfield  {journal} {\bibinfo  {journal}
  {Journal of Retailing and Consumer Services}\ }\textbf {\bibinfo {volume}
  {18}},\ \bibinfo {pages} {348} (\bibinfo {year} {2011})}\BibitemShut
  {NoStop}%
\bibitem [{\citenamefont {Acemo{\u{g}}lu}\ \emph {et~al.}(2013)\citenamefont
  {Acemo{\u{g}}lu}, \citenamefont {Como}, \citenamefont {Fagnani},\ and\
  \citenamefont {Ozdaglar}}]{acemouglu2013opinion}%
  \BibitemOpen
  \bibfield  {author} {\bibinfo {author} {\bibfnamefont {D.}~\bibnamefont
  {Acemo{\u{g}}lu}}, \bibinfo {author} {\bibfnamefont {G.}~\bibnamefont
  {Como}}, \bibinfo {author} {\bibfnamefont {F.}~\bibnamefont {Fagnani}}, \
  and\ \bibinfo {author} {\bibfnamefont {A.}~\bibnamefont {Ozdaglar}},\
  }\href@noop {} {\bibfield  {journal} {\bibinfo  {journal} {Mathematics of
  Operations Research}\ }\textbf {\bibinfo {volume} {38}},\ \bibinfo {pages}
  {1} (\bibinfo {year} {2013})}\BibitemShut {NoStop}%
\bibitem [{\citenamefont {Gomes}\ \emph {et~al.}(2019)\citenamefont {Gomes},
  \citenamefont {Reia}, \citenamefont {Rodrigues},\ and\ \citenamefont
  {Fontanari}}]{gomes2019mobility}%
  \BibitemOpen
  \bibfield  {author} {\bibinfo {author} {\bibfnamefont {P.~F.}\ \bibnamefont
  {Gomes}}, \bibinfo {author} {\bibfnamefont {S.~M.}\ \bibnamefont {Reia}},
  \bibinfo {author} {\bibfnamefont {F.~A.}\ \bibnamefont {Rodrigues}}, \ and\
  \bibinfo {author} {\bibfnamefont {J.~F.}\ \bibnamefont {Fontanari}},\
  }\href@noop {} {\bibfield  {journal} {\bibinfo  {journal} {Physical Review
  E}\ }\textbf {\bibinfo {volume} {99}},\ \bibinfo {pages} {032301} (\bibinfo
  {year} {2019})}\BibitemShut {NoStop}%
\bibitem [{\citenamefont {Gracia-L{\'a}zaro}\ \emph {et~al.}(2011)\citenamefont
  {Gracia-L{\'a}zaro}, \citenamefont {Quijandr{\'\i}a}, \citenamefont
  {Hern{\'a}ndez}, \citenamefont {Flor{\'\i}a},\ and\ \citenamefont
  {Moreno}}]{gracia2011coevolutionary}%
  \BibitemOpen
  \bibfield  {author} {\bibinfo {author} {\bibfnamefont {C.}~\bibnamefont
  {Gracia-L{\'a}zaro}}, \bibinfo {author} {\bibfnamefont {F.}~\bibnamefont
  {Quijandr{\'\i}a}}, \bibinfo {author} {\bibfnamefont {L.}~\bibnamefont
  {Hern{\'a}ndez}}, \bibinfo {author} {\bibfnamefont {L.~M.}\ \bibnamefont
  {Flor{\'\i}a}}, \ and\ \bibinfo {author} {\bibfnamefont {Y.}~\bibnamefont
  {Moreno}},\ }\href@noop {} {\bibfield  {journal} {\bibinfo  {journal}
  {Physical Review E}\ }\textbf {\bibinfo {volume} {84}},\ \bibinfo {pages}
  {067101} (\bibinfo {year} {2011})}\BibitemShut {NoStop}%
\bibitem [{\citenamefont {Sznajd-Weron}\ and\ \citenamefont
  {Sznajd}(2000)}]{sznajd2000opinion}%
  \BibitemOpen
  \bibfield  {author} {\bibinfo {author} {\bibfnamefont {K.}~\bibnamefont
  {Sznajd-Weron}}\ and\ \bibinfo {author} {\bibfnamefont {J.}~\bibnamefont
  {Sznajd}},\ }\href@noop {} {\bibfield  {journal} {\bibinfo  {journal}
  {International Journal of Modern Physics C}\ }\textbf {\bibinfo {volume}
  {11}},\ \bibinfo {pages} {1157} (\bibinfo {year} {2000})}\BibitemShut
  {NoStop}%
\bibitem [{\citenamefont {Lee}\ \emph {et~al.}(2019)\citenamefont {Lee},
  \citenamefont {Karimi}, \citenamefont {Wagner}, \citenamefont {Jo},
  \citenamefont {Strohmaier},\ and\ \citenamefont
  {Galesic}}]{lee2019homophily}%
  \BibitemOpen
  \bibfield  {author} {\bibinfo {author} {\bibfnamefont {E.}~\bibnamefont
  {Lee}}, \bibinfo {author} {\bibfnamefont {F.}~\bibnamefont {Karimi}},
  \bibinfo {author} {\bibfnamefont {C.}~\bibnamefont {Wagner}}, \bibinfo
  {author} {\bibfnamefont {H.-H.}\ \bibnamefont {Jo}}, \bibinfo {author}
  {\bibfnamefont {M.}~\bibnamefont {Strohmaier}}, \ and\ \bibinfo {author}
  {\bibfnamefont {M.}~\bibnamefont {Galesic}},\ }\href@noop {} {\bibfield
  {journal} {\bibinfo  {journal} {Nature human behaviour}\ }\textbf {\bibinfo
  {volume} {3}},\ \bibinfo {pages} {1078} (\bibinfo {year} {2019})}\BibitemShut
  {NoStop}%
\bibitem [{\citenamefont {Benatti}\ \emph {et~al.}(2020)\citenamefont
  {Benatti}, \citenamefont {de~Arruda}, \citenamefont {Silva}, \citenamefont
  {Comin},\ and\ \citenamefont {da~Fontoura~Costa}}]{benatti2019opinion}%
  \BibitemOpen
  \bibfield  {author} {\bibinfo {author} {\bibfnamefont {A.}~\bibnamefont
  {Benatti}}, \bibinfo {author} {\bibfnamefont {H.~F.}\ \bibnamefont
  {de~Arruda}}, \bibinfo {author} {\bibfnamefont {F.~N.}\ \bibnamefont
  {Silva}}, \bibinfo {author} {\bibfnamefont {C.~H.}\ \bibnamefont {Comin}}, \
  and\ \bibinfo {author} {\bibfnamefont {L.}~\bibnamefont
  {da~Fontoura~Costa}},\ }\href@noop {} {\bibfield  {journal} {\bibinfo
  {journal} {Journal of Statistical Mechanics: Theory and Experiment}\ }\textbf
  {\bibinfo {volume} {2020}},\ \bibinfo {pages} {023407} (\bibinfo {year}
  {2020})}\BibitemShut {NoStop}%
\bibitem [{\citenamefont {He}\ \emph {et~al.}(2004)\citenamefont {He},
  \citenamefont {Li},\ and\ \citenamefont {Luo}}]{he2004sznajd}%
  \BibitemOpen
  \bibfield  {author} {\bibinfo {author} {\bibfnamefont {M.}~\bibnamefont
  {He}}, \bibinfo {author} {\bibfnamefont {B.}~\bibnamefont {Li}}, \ and\
  \bibinfo {author} {\bibfnamefont {L.}~\bibnamefont {Luo}},\ }\href@noop {}
  {\bibfield  {journal} {\bibinfo  {journal} {International Journal of Modern
  Physics C}\ }\textbf {\bibinfo {volume} {15}},\ \bibinfo {pages} {997}
  (\bibinfo {year} {2004})}\BibitemShut {NoStop}%
\bibitem [{\citenamefont {Holme}\ and\ \citenamefont
  {Newman}(2006)}]{holme2006nonequilibrium}%
  \BibitemOpen
  \bibfield  {author} {\bibinfo {author} {\bibfnamefont {P.}~\bibnamefont
  {Holme}}\ and\ \bibinfo {author} {\bibfnamefont {M.~E.}\ \bibnamefont
  {Newman}},\ }\href@noop {} {\bibfield  {journal} {\bibinfo  {journal}
  {Physical Review E}\ }\textbf {\bibinfo {volume} {74}},\ \bibinfo {pages}
  {056108} (\bibinfo {year} {2006})}\BibitemShut {NoStop}%
\bibitem [{\citenamefont {Fu}\ and\ \citenamefont
  {Wang}(2008)}]{fu2008coevolutionary}%
  \BibitemOpen
  \bibfield  {author} {\bibinfo {author} {\bibfnamefont {F.}~\bibnamefont
  {Fu}}\ and\ \bibinfo {author} {\bibfnamefont {L.}~\bibnamefont {Wang}},\
  }\href@noop {} {\bibfield  {journal} {\bibinfo  {journal} {Physical Review
  E}\ }\textbf {\bibinfo {volume} {78}},\ \bibinfo {pages} {016104} (\bibinfo
  {year} {2008})}\BibitemShut {NoStop}%
\bibitem [{\citenamefont {Durrett}\ \emph {et~al.}(2012)\citenamefont
  {Durrett}, \citenamefont {Gleeson}, \citenamefont {Lloyd}, \citenamefont
  {Mucha}, \citenamefont {Shi}, \citenamefont {Sivakoff}, \citenamefont
  {Socolar},\ and\ \citenamefont {Varghese}}]{durrett2012graph}%
  \BibitemOpen
  \bibfield  {author} {\bibinfo {author} {\bibfnamefont {R.}~\bibnamefont
  {Durrett}}, \bibinfo {author} {\bibfnamefont {J.~P.}\ \bibnamefont
  {Gleeson}}, \bibinfo {author} {\bibfnamefont {A.~L.}\ \bibnamefont {Lloyd}},
  \bibinfo {author} {\bibfnamefont {P.~J.}\ \bibnamefont {Mucha}}, \bibinfo
  {author} {\bibfnamefont {F.}~\bibnamefont {Shi}}, \bibinfo {author}
  {\bibfnamefont {D.}~\bibnamefont {Sivakoff}}, \bibinfo {author}
  {\bibfnamefont {J.~E.}\ \bibnamefont {Socolar}}, \ and\ \bibinfo {author}
  {\bibfnamefont {C.}~\bibnamefont {Varghese}},\ }\href@noop {} {\bibfield
  {journal} {\bibinfo  {journal} {Proceedings of the National Academy of
  Sciences}\ }\textbf {\bibinfo {volume} {109}},\ \bibinfo {pages} {3682}
  (\bibinfo {year} {2012})}\BibitemShut {NoStop}%
\bibitem [{\citenamefont {Iniguez}\ \emph {et~al.}(2009)\citenamefont
  {Iniguez}, \citenamefont {Kert{\'e}sz}, \citenamefont {Kaski},\ and\
  \citenamefont {Barrio}}]{iniguez2009opinion}%
  \BibitemOpen
  \bibfield  {author} {\bibinfo {author} {\bibfnamefont {G.}~\bibnamefont
  {Iniguez}}, \bibinfo {author} {\bibfnamefont {J.}~\bibnamefont
  {Kert{\'e}sz}}, \bibinfo {author} {\bibfnamefont {K.~K.}\ \bibnamefont
  {Kaski}}, \ and\ \bibinfo {author} {\bibfnamefont {R.~A.}\ \bibnamefont
  {Barrio}},\ }\href@noop {} {\bibfield  {journal} {\bibinfo  {journal}
  {Physical Review E}\ }\textbf {\bibinfo {volume} {80}},\ \bibinfo {pages}
  {066119} (\bibinfo {year} {2009})}\BibitemShut {NoStop}%
\bibitem [{\citenamefont {Del~Vicario}\ \emph {et~al.}(2015)\citenamefont
  {Del~Vicario}, \citenamefont {Bessi}, \citenamefont {Zollo}, \citenamefont
  {Petroni}, \citenamefont {Scala}, \citenamefont {Caldarelli}, \citenamefont
  {Stanley},\ and\ \citenamefont {Quattrociocchi}}]{del2015echo}%
  \BibitemOpen
  \bibfield  {author} {\bibinfo {author} {\bibfnamefont {M.}~\bibnamefont
  {Del~Vicario}}, \bibinfo {author} {\bibfnamefont {A.}~\bibnamefont {Bessi}},
  \bibinfo {author} {\bibfnamefont {F.}~\bibnamefont {Zollo}}, \bibinfo
  {author} {\bibfnamefont {F.}~\bibnamefont {Petroni}}, \bibinfo {author}
  {\bibfnamefont {A.}~\bibnamefont {Scala}}, \bibinfo {author} {\bibfnamefont
  {G.}~\bibnamefont {Caldarelli}}, \bibinfo {author} {\bibfnamefont {H.~E.}\
  \bibnamefont {Stanley}}, \ and\ \bibinfo {author} {\bibfnamefont
  {W.}~\bibnamefont {Quattrociocchi}},\ }\href@noop {} {\bibfield  {journal}
  {\bibinfo  {journal} {arXiv preprint arXiv:1509.00189}\ } (\bibinfo {year}
  {2015})}\BibitemShut {NoStop}%
\bibitem [{\citenamefont {T{\"o}rnberg}(2018)}]{tornberg2018echo}%
  \BibitemOpen
  \bibfield  {author} {\bibinfo {author} {\bibfnamefont {P.}~\bibnamefont
  {T{\"o}rnberg}},\ }\href@noop {} {\bibfield  {journal} {\bibinfo  {journal}
  {PloS one}\ }\textbf {\bibinfo {volume} {13}},\ \bibinfo {pages} {e0203958}
  (\bibinfo {year} {2018})}\BibitemShut {NoStop}%
\bibitem [{\citenamefont {Jasny}\ \emph {et~al.}(2015)\citenamefont {Jasny},
  \citenamefont {Waggle},\ and\ \citenamefont {Fisher}}]{jasny2015empirical}%
  \BibitemOpen
  \bibfield  {author} {\bibinfo {author} {\bibfnamefont {L.}~\bibnamefont
  {Jasny}}, \bibinfo {author} {\bibfnamefont {J.}~\bibnamefont {Waggle}}, \
  and\ \bibinfo {author} {\bibfnamefont {D.~R.}\ \bibnamefont {Fisher}},\
  }\href@noop {} {\bibfield  {journal} {\bibinfo  {journal} {Nature Climate
  Change}\ }\textbf {\bibinfo {volume} {5}},\ \bibinfo {pages} {782} (\bibinfo
  {year} {2015})}\BibitemShut {NoStop}%
\bibitem [{\citenamefont {Jasny}\ \emph {et~al.}(2018)\citenamefont {Jasny},
  \citenamefont {Dewey}, \citenamefont {Robertson}, \citenamefont {Yagatich},
  \citenamefont {Dubin}, \citenamefont {Waggle},\ and\ \citenamefont
  {Fisher}}]{jasny2018shifting}%
  \BibitemOpen
  \bibfield  {author} {\bibinfo {author} {\bibfnamefont {L.}~\bibnamefont
  {Jasny}}, \bibinfo {author} {\bibfnamefont {A.~M.}\ \bibnamefont {Dewey}},
  \bibinfo {author} {\bibfnamefont {A.~G.}\ \bibnamefont {Robertson}}, \bibinfo
  {author} {\bibfnamefont {W.}~\bibnamefont {Yagatich}}, \bibinfo {author}
  {\bibfnamefont {A.~H.}\ \bibnamefont {Dubin}}, \bibinfo {author}
  {\bibfnamefont {J.~M.}\ \bibnamefont {Waggle}}, \ and\ \bibinfo {author}
  {\bibfnamefont {D.~R.}\ \bibnamefont {Fisher}},\ }\href@noop {} {\bibfield
  {journal} {\bibinfo  {journal} {PloS one}\ }\textbf {\bibinfo {volume}
  {13}},\ \bibinfo {pages} {e0203463} (\bibinfo {year} {2018})}\BibitemShut
  {NoStop}%
\bibitem [{\citenamefont {Quattrociocchi}\ \emph {et~al.}(2016)\citenamefont
  {Quattrociocchi}, \citenamefont {Scala},\ and\ \citenamefont
  {Sunstein}}]{quattrociocchi2016echo}%
  \BibitemOpen
  \bibfield  {author} {\bibinfo {author} {\bibfnamefont {W.}~\bibnamefont
  {Quattrociocchi}}, \bibinfo {author} {\bibfnamefont {A.}~\bibnamefont
  {Scala}}, \ and\ \bibinfo {author} {\bibfnamefont {C.~R.}\ \bibnamefont
  {Sunstein}},\ }\href@noop {} {\bibfield  {journal} {\bibinfo  {journal}
  {Available at SSRN 2795110}\ } (\bibinfo {year} {2016})}\BibitemShut
  {NoStop}%
\bibitem [{\citenamefont {Del~Vicario}\ \emph {et~al.}(2016)\citenamefont
  {Del~Vicario}, \citenamefont {Bessi}, \citenamefont {Zollo}, \citenamefont
  {Petroni}, \citenamefont {Scala}, \citenamefont {Caldarelli}, \citenamefont
  {Stanley},\ and\ \citenamefont {Quattrociocchi}}]{del2016spreading}%
  \BibitemOpen
  \bibfield  {author} {\bibinfo {author} {\bibfnamefont {M.}~\bibnamefont
  {Del~Vicario}}, \bibinfo {author} {\bibfnamefont {A.}~\bibnamefont {Bessi}},
  \bibinfo {author} {\bibfnamefont {F.}~\bibnamefont {Zollo}}, \bibinfo
  {author} {\bibfnamefont {F.}~\bibnamefont {Petroni}}, \bibinfo {author}
  {\bibfnamefont {A.}~\bibnamefont {Scala}}, \bibinfo {author} {\bibfnamefont
  {G.}~\bibnamefont {Caldarelli}}, \bibinfo {author} {\bibfnamefont {H.~E.}\
  \bibnamefont {Stanley}}, \ and\ \bibinfo {author} {\bibfnamefont
  {W.}~\bibnamefont {Quattrociocchi}},\ }\href@noop {} {\bibfield  {journal}
  {\bibinfo  {journal} {Proceedings of the National Academy of Sciences}\
  }\textbf {\bibinfo {volume} {113}},\ \bibinfo {pages} {554} (\bibinfo {year}
  {2016})}\BibitemShut {NoStop}%
\bibitem [{\citenamefont {Cinelli}\ \emph {et~al.}(2020)\citenamefont
  {Cinelli}, \citenamefont {Morales}, \citenamefont {Galeazzi}, \citenamefont
  {Quattrociocchi},\ and\ \citenamefont {Starnini}}]{cinelli2020echo}%
  \BibitemOpen
  \bibfield  {author} {\bibinfo {author} {\bibfnamefont {M.}~\bibnamefont
  {Cinelli}}, \bibinfo {author} {\bibfnamefont {G.~D.~F.}\ \bibnamefont
  {Morales}}, \bibinfo {author} {\bibfnamefont {A.}~\bibnamefont {Galeazzi}},
  \bibinfo {author} {\bibfnamefont {W.}~\bibnamefont {Quattrociocchi}}, \ and\
  \bibinfo {author} {\bibfnamefont {M.}~\bibnamefont {Starnini}},\ }\href@noop
  {} {\bibfield  {journal} {\bibinfo  {journal} {arXiv preprint
  arXiv:2004.09603}\ } (\bibinfo {year} {2020})}\BibitemShut {NoStop}%
\bibitem [{\citenamefont {Dong}\ \emph {et~al.}(2018)\citenamefont {Dong},
  \citenamefont {Zhan}, \citenamefont {Kou}, \citenamefont {Ding},\ and\
  \citenamefont {Liang}}]{dong2018survey}%
  \BibitemOpen
  \bibfield  {author} {\bibinfo {author} {\bibfnamefont {Y.}~\bibnamefont
  {Dong}}, \bibinfo {author} {\bibfnamefont {M.}~\bibnamefont {Zhan}}, \bibinfo
  {author} {\bibfnamefont {G.}~\bibnamefont {Kou}}, \bibinfo {author}
  {\bibfnamefont {Z.}~\bibnamefont {Ding}}, \ and\ \bibinfo {author}
  {\bibfnamefont {H.}~\bibnamefont {Liang}},\ }\href@noop {} {\bibfield
  {journal} {\bibinfo  {journal} {Information Fusion}\ }\textbf {\bibinfo
  {volume} {43}},\ \bibinfo {pages} {57} (\bibinfo {year} {2018})}\BibitemShut
  {NoStop}%
\bibitem [{\citenamefont {Crokidakis}\ \emph {et~al.}(2014)\citenamefont
  {Crokidakis}, \citenamefont {Blanco},\ and\ \citenamefont
  {Anteneodo}}]{crokidakis2014impact}%
  \BibitemOpen
  \bibfield  {author} {\bibinfo {author} {\bibfnamefont {N.}~\bibnamefont
  {Crokidakis}}, \bibinfo {author} {\bibfnamefont {V.~H.}\ \bibnamefont
  {Blanco}}, \ and\ \bibinfo {author} {\bibfnamefont {C.}~\bibnamefont
  {Anteneodo}},\ }\href@noop {} {\bibfield  {journal} {\bibinfo  {journal}
  {Physical Review E}\ }\textbf {\bibinfo {volume} {89}},\ \bibinfo {pages}
  {013310} (\bibinfo {year} {2014})}\BibitemShut {NoStop}%
\bibitem [{\citenamefont {Galam}\ and\ \citenamefont
  {Jacobs}(2007)}]{galam2007role}%
  \BibitemOpen
  \bibfield  {author} {\bibinfo {author} {\bibfnamefont {S.}~\bibnamefont
  {Galam}}\ and\ \bibinfo {author} {\bibfnamefont {F.}~\bibnamefont {Jacobs}},\
  }\href@noop {} {\bibfield  {journal} {\bibinfo  {journal} {Physica A:
  Statistical Mechanics and its Applications}\ }\textbf {\bibinfo {volume}
  {381}},\ \bibinfo {pages} {366} (\bibinfo {year} {2007})}\BibitemShut
  {NoStop}%
\bibitem [{\citenamefont {Vandello}\ \emph {et~al.}(2007)\citenamefont
  {Vandello}, \citenamefont {Goldschmied},\ and\ \citenamefont
  {Richards}}]{vandello2007appeal}%
  \BibitemOpen
  \bibfield  {author} {\bibinfo {author} {\bibfnamefont {J.~A.}\ \bibnamefont
  {Vandello}}, \bibinfo {author} {\bibfnamefont {N.~P.}\ \bibnamefont
  {Goldschmied}}, \ and\ \bibinfo {author} {\bibfnamefont {D.~A.}\ \bibnamefont
  {Richards}},\ }\href@noop {} {\bibfield  {journal} {\bibinfo  {journal}
  {Personality and Social Psychology Bulletin}\ }\textbf {\bibinfo {volume}
  {33}},\ \bibinfo {pages} {1603} (\bibinfo {year} {2007})}\BibitemShut
  {NoStop}%
\bibitem [{\citenamefont {Ulmer}(1978)}]{ulmer1978selecting}%
  \BibitemOpen
  \bibfield  {author} {\bibinfo {author} {\bibfnamefont {S.~S.}\ \bibnamefont
  {Ulmer}},\ }\href@noop {} {\bibfield  {journal} {\bibinfo  {journal} {The
  American Political Science Review}\ ,\ \bibinfo {pages} {902}} (\bibinfo
  {year} {1978})}\BibitemShut {NoStop}%
\bibitem [{\citenamefont {Frazier}\ and\ \citenamefont
  {Snyder}(1991)}]{frazier1991underdog}%
  \BibitemOpen
  \bibfield  {author} {\bibinfo {author} {\bibfnamefont {J.~A.}\ \bibnamefont
  {Frazier}}\ and\ \bibinfo {author} {\bibfnamefont {E.~E.}\ \bibnamefont
  {Snyder}},\ }\href@noop {} {\bibfield  {journal} {\bibinfo  {journal}
  {Sociology of Sport Journal}\ }\textbf {\bibinfo {volume} {8}},\ \bibinfo
  {pages} {380} (\bibinfo {year} {1991})}\BibitemShut {NoStop}%
\bibitem [{\citenamefont {Watts}\ and\ \citenamefont
  {Strogatz}(1998{\natexlab{a}})}]{watts1998collective}%
  \BibitemOpen
  \bibfield  {author} {\bibinfo {author} {\bibfnamefont {D.~J.}\ \bibnamefont
  {Watts}}\ and\ \bibinfo {author} {\bibfnamefont {S.~H.}\ \bibnamefont
  {Strogatz}},\ }\href@noop {} {\bibfield  {journal} {\bibinfo  {journal}
  {Nature}\ }\textbf {\bibinfo {volume} {393}},\ \bibinfo {pages} {440}
  (\bibinfo {year} {1998}{\natexlab{a}})}\BibitemShut {NoStop}%
\bibitem [{\citenamefont {Newman}(2006)}]{newman2006modularity}%
  \BibitemOpen
  \bibfield  {author} {\bibinfo {author} {\bibfnamefont {M.~E.}\ \bibnamefont
  {Newman}},\ }\href@noop {} {\bibfield  {journal} {\bibinfo  {journal}
  {Proceedings of the national academy of sciences}\ }\textbf {\bibinfo
  {volume} {103}},\ \bibinfo {pages} {8577} (\bibinfo {year}
  {2006})}\BibitemShut {NoStop}%
\bibitem [{\citenamefont {Jost}(2006)}]{jost2006entropy}%
  \BibitemOpen
  \bibfield  {author} {\bibinfo {author} {\bibfnamefont {L.}~\bibnamefont
  {Jost}},\ }\href@noop {} {\bibfield  {journal} {\bibinfo  {journal} {Oikos}\
  }\textbf {\bibinfo {volume} {113}},\ \bibinfo {pages} {363} (\bibinfo {year}
  {2006})}\BibitemShut {NoStop}%
\bibitem [{\citenamefont {Pielou}(1966)}]{pielou1966shannon}%
  \BibitemOpen
  \bibfield  {author} {\bibinfo {author} {\bibfnamefont {E.~C.}\ \bibnamefont
  {Pielou}},\ }\href@noop {} {\bibfield  {journal} {\bibinfo  {journal} {The
  American Naturalist}\ }\textbf {\bibinfo {volume} {100}},\ \bibinfo {pages}
  {463} (\bibinfo {year} {1966})}\BibitemShut {NoStop}%
\bibitem [{\citenamefont {Hill}(1973)}]{hill1973diversity}%
  \BibitemOpen
  \bibfield  {author} {\bibinfo {author} {\bibfnamefont {M.~O.}\ \bibnamefont
  {Hill}},\ }\href@noop {} {\bibfield  {journal} {\bibinfo  {journal}
  {Ecology}\ }\textbf {\bibinfo {volume} {54}},\ \bibinfo {pages} {427}
  (\bibinfo {year} {1973})}\BibitemShut {NoStop}%
\bibitem [{\citenamefont {Chao}\ \emph {et~al.}(2016)\citenamefont {Chao},
  \citenamefont {Chiu},\ and\ \citenamefont {Jost}}]{chao2016phylogenetic}%
  \BibitemOpen
  \bibfield  {author} {\bibinfo {author} {\bibfnamefont {A.}~\bibnamefont
  {Chao}}, \bibinfo {author} {\bibfnamefont {C.-H.}\ \bibnamefont {Chiu}}, \
  and\ \bibinfo {author} {\bibfnamefont {L.}~\bibnamefont {Jost}},\ }\href@noop
  {} {\bibfield  {journal} {\bibinfo  {journal} {Biodiversity Conservation and
  Phylogenetic Systematics}\ ,\ \bibinfo {pages} {141}} (\bibinfo {year}
  {2016})}\BibitemShut {NoStop}%
\bibitem [{\citenamefont {Messias}\ \emph {et~al.}(2018)\citenamefont
  {Messias}, \citenamefont {Silva}, \citenamefont {Comin},\ and\ \citenamefont
  {da~F~Costa}}]{messias2018can}%
  \BibitemOpen
  \bibfield  {author} {\bibinfo {author} {\bibfnamefont {B.}~\bibnamefont
  {Messias}}, \bibinfo {author} {\bibfnamefont {F.~N.}\ \bibnamefont {Silva}},
  \bibinfo {author} {\bibfnamefont {C.~H.}\ \bibnamefont {Comin}}, \ and\
  \bibinfo {author} {\bibfnamefont {L.}~\bibnamefont {da~F~Costa}},\
  }\href@noop {} {\bibfield  {journal} {\bibinfo  {journal} {arXiv preprint
  arXiv:1809.00729}\ } (\bibinfo {year} {2018})}\BibitemShut {NoStop}%
\bibitem [{\citenamefont {Watts}\ and\ \citenamefont
  {Strogatz}(1998{\natexlab{b}})}]{watts1998WS}%
  \BibitemOpen
  \bibfield  {author} {\bibinfo {author} {\bibfnamefont {D.~J.}\ \bibnamefont
  {Watts}}\ and\ \bibinfo {author} {\bibfnamefont {S.~H.}\ \bibnamefont
  {Strogatz}},\ }\href@noop {} {\bibfield  {journal} {\bibinfo  {journal}
  {nature}\ }\textbf {\bibinfo {volume} {393}},\ \bibinfo {pages} {440}
  (\bibinfo {year} {1998}{\natexlab{b}})}\BibitemShut {NoStop}%
\bibitem [{\citenamefont {Erd\H{o}s}\ and\ \citenamefont
  {A.}(1959)}]{erdos1959random}%
  \BibitemOpen
  \bibfield  {author} {\bibinfo {author} {\bibfnamefont {P.}~\bibnamefont
  {Erd\H{o}s}}\ and\ \bibinfo {author} {\bibfnamefont {R.}~\bibnamefont {A.}},\
  }\href@noop {} {\bibfield  {journal} {\bibinfo  {journal} {Publ. Math.
  (Debrecen)}\ }\textbf {\bibinfo {volume} {6}},\ \bibinfo {pages} {290}
  (\bibinfo {year} {1959})}\BibitemShut {NoStop}%
\bibitem [{\citenamefont {Barab{\'a}si}\ and\ \citenamefont
  {Albert}(1999)}]{barabasi1999BA}%
  \BibitemOpen
  \bibfield  {author} {\bibinfo {author} {\bibfnamefont {A.-L.}\ \bibnamefont
  {Barab{\'a}si}}\ and\ \bibinfo {author} {\bibfnamefont {R.}~\bibnamefont
  {Albert}},\ }\href@noop {} {\bibfield  {journal} {\bibinfo  {journal}
  {science}\ }\textbf {\bibinfo {volume} {286}},\ \bibinfo {pages} {509}
  (\bibinfo {year} {1999})}\BibitemShut {NoStop}%
\bibitem [{\citenamefont {Penrose}(2003)}]{penrose2003random}%
  \BibitemOpen
  \bibfield  {author} {\bibinfo {author} {\bibfnamefont {M.}~\bibnamefont
  {Penrose}},\ }\href@noop {} {\emph {\bibinfo {title} {Random geometric
  graphs}}},\ \bibinfo {number} {5}\ (\bibinfo  {publisher} {Oxford University
  Press},\ \bibinfo {year} {2003})\BibitemShut {NoStop}%
\bibitem [{\citenamefont {Holland}\ \emph {et~al.}(1983)\citenamefont
  {Holland}, \citenamefont {Laskey},\ and\ \citenamefont
  {Leinhardt}}]{holland1983stochastic}%
  \BibitemOpen
  \bibfield  {author} {\bibinfo {author} {\bibfnamefont {P.~W.}\ \bibnamefont
  {Holland}}, \bibinfo {author} {\bibfnamefont {K.~B.}\ \bibnamefont {Laskey}},
  \ and\ \bibinfo {author} {\bibfnamefont {S.}~\bibnamefont {Leinhardt}},\
  }\href@noop {} {\bibfield  {journal} {\bibinfo  {journal} {Social networks}\
  }\textbf {\bibinfo {volume} {5}},\ \bibinfo {pages} {109} (\bibinfo {year}
  {1983})}\BibitemShut {NoStop}%
\bibitem [{\citenamefont {da~F~Costa}\ \emph {et~al.}(2007)\citenamefont
  {da~F~Costa}, \citenamefont {Rodrigues}, \citenamefont {Travieso},\ and\
  \citenamefont {Villas~Boas}}]{costa2007characterization}%
  \BibitemOpen
  \bibfield  {author} {\bibinfo {author} {\bibfnamefont {L.}~\bibnamefont
  {da~F~Costa}}, \bibinfo {author} {\bibfnamefont {F.~A.}\ \bibnamefont
  {Rodrigues}}, \bibinfo {author} {\bibfnamefont {G.}~\bibnamefont {Travieso}},
  \ and\ \bibinfo {author} {\bibfnamefont {P.~R.}\ \bibnamefont
  {Villas~Boas}},\ }\href@noop {} {\bibfield  {journal} {\bibinfo  {journal}
  {Advances in physics}\ }\textbf {\bibinfo {volume} {56}},\ \bibinfo {pages}
  {167} (\bibinfo {year} {2007})}\BibitemShut {NoStop}%
\bibitem [{\citenamefont {Jolliffe}(2011)}]{jolliffe2011principal}%
  \BibitemOpen
  \bibfield  {author} {\bibinfo {author} {\bibfnamefont {I.}~\bibnamefont
  {Jolliffe}},\ }\href@noop {} {\emph {\bibinfo {title} {Principal component
  analysis}}}\ (\bibinfo  {publisher} {Springer},\ \bibinfo {year}
  {2011})\BibitemShut {NoStop}%
\bibitem [{\citenamefont {Gewers}\ \emph {et~al.}(2018)\citenamefont {Gewers},
  \citenamefont {Ferreira}, \citenamefont {de~Arruda}, \citenamefont {Silva},
  \citenamefont {Comin}, \citenamefont {Amancio},\ and\ \citenamefont
  {da~F~Costa}}]{gewers2018principal}%
  \BibitemOpen
  \bibfield  {author} {\bibinfo {author} {\bibfnamefont {F.~L.}\ \bibnamefont
  {Gewers}}, \bibinfo {author} {\bibfnamefont {G.~R.}\ \bibnamefont
  {Ferreira}}, \bibinfo {author} {\bibfnamefont {H.~F.}\ \bibnamefont
  {de~Arruda}}, \bibinfo {author} {\bibfnamefont {F.~N.}\ \bibnamefont
  {Silva}}, \bibinfo {author} {\bibfnamefont {C.~H.}\ \bibnamefont {Comin}},
  \bibinfo {author} {\bibfnamefont {D.~R.}\ \bibnamefont {Amancio}}, \ and\
  \bibinfo {author} {\bibfnamefont {L.}~\bibnamefont {da~F~Costa}},\
  }\href@noop {} {\bibfield  {journal} {\bibinfo  {journal} {arXiv preprint
  arXiv:1804.02502}\ } (\bibinfo {year} {2018})}\BibitemShut {NoStop}%
\bibitem [{\citenamefont {Fortunato}(2010)}]{fortunato2010community}%
  \BibitemOpen
  \bibfield  {author} {\bibinfo {author} {\bibfnamefont {S.}~\bibnamefont
  {Fortunato}},\ }\href@noop {} {\bibfield  {journal} {\bibinfo  {journal}
  {Physics reports}\ }\textbf {\bibinfo {volume} {486}},\ \bibinfo {pages} {75}
  (\bibinfo {year} {2010})}\BibitemShut {NoStop}%
\bibitem [{\citenamefont {Fortunato}\ and\ \citenamefont
  {Barthelemy}(2007)}]{fortunato2007resolution}%
  \BibitemOpen
  \bibfield  {author} {\bibinfo {author} {\bibfnamefont {S.}~\bibnamefont
  {Fortunato}}\ and\ \bibinfo {author} {\bibfnamefont {M.}~\bibnamefont
  {Barthelemy}},\ }\href@noop {} {\bibfield  {journal} {\bibinfo  {journal}
  {Proceedings of the national academy of sciences}\ }\textbf {\bibinfo
  {volume} {104}},\ \bibinfo {pages} {36} (\bibinfo {year} {2007})}\BibitemShut
  {NoStop}%
\bibitem [{\citenamefont {Silva}\ \emph {et~al.}(2016)\citenamefont {Silva},
  \citenamefont {Amancio}, \citenamefont {Bardosova}, \citenamefont
  {da~F~Costa},\ and\ \citenamefont {Oliveira~Jr}}]{silva2016using}%
  \BibitemOpen
  \bibfield  {author} {\bibinfo {author} {\bibfnamefont {F.~N.}\ \bibnamefont
  {Silva}}, \bibinfo {author} {\bibfnamefont {D.~R.}\ \bibnamefont {Amancio}},
  \bibinfo {author} {\bibfnamefont {M.}~\bibnamefont {Bardosova}}, \bibinfo
  {author} {\bibfnamefont {L.}~\bibnamefont {da~F~Costa}}, \ and\ \bibinfo
  {author} {\bibfnamefont {O.~N.}\ \bibnamefont {Oliveira~Jr}},\ }\href@noop {}
  {\bibfield  {journal} {\bibinfo  {journal} {Journal of Informetrics}\
  }\textbf {\bibinfo {volume} {10}},\ \bibinfo {pages} {487} (\bibinfo {year}
  {2016})}\BibitemShut {NoStop}%
\end{thebibliography}%

\end{document}